\documentclass[journal]{IEEEtran}

\usepackage{graphicx}
\usepackage{amssymb, amsmath, bm, mathtools,array}
\usepackage{textcomp}
\usepackage{hyperref}
\usepackage{verbatim}
\usepackage{lipsum} 
\usepackage{multirow}
\usepackage{siunitx}
\usepackage{diagbox}
\usepackage{booktabs}

\newcommand{\bs}[1]{\boldsymbol{#1}}

\newcommand{\ceil}[1]{\left\lceil #1 \right\rceil}

\RequirePackage{color}\definecolor{BLUE}{rgb}{0,0,1} 

\ifCLASSINFOpdf
  \graphicspath{{./}}
\else
\fi

\ifCLASSOPTIONcompsoc
  \usepackage[caption=false,font=normalsize,labelfont=sf,textfont=sf]{subfig}
\else
  \usepackage[caption=false,font=footnotesize]{subfig}
\fi

\begin{document}
\title{{Neural source-filter waveform models for statistical parametric speech synthesis}}
\author{Xin~Wang,~\IEEEmembership{Member,~IEEE,}
        Shinji~Takaki,~\IEEEmembership{Member,~IEEE,}
        Junichi~Yamagishi,~\IEEEmembership{Senior~Member,~IEEE}
\thanks{Xin Wang is with National Institute of Informatics, Tokyo, 101-8340, Japan. e-mail:wangxin@nii.ac.jp}
\thanks{Shinji Takaki is with the Nagoya Institute of Technology, Japan. e-mail: takaki@sp.nitech.ac.jp}
\thanks{Junichi Yamagishi is with the National Institute of Informatics, Tokyo, 101-8340, Japan and also with Centre for Speech Technology Research, University of Edinburgh, UK. e-mail: jyamagis@nii.ac.jp.}
\thanks{The guest editor coordinating the review of this manuscript was Dr. Heiga Zen. This work was partially supported by JST CREST Grant Number JPMJCR18A6, Japan and by MEXT KAKENHI Grant Numbers (19K24371,16H06302, 16K16096, 17H04687, 18H04120, 18H04112, 18KT0051), Japan.}
}

\markboth{Manuscript}%
{Shell \MakeLowercase{\textit{et al.}}: Bare Demo of IEEEtran.cls for IEEE Journals}

\maketitle

\begin{abstract}
Neural waveform models have demonstrated better performance than conventional vocoders for statistical parametric speech synthesis. 
One of the best models, called WaveNet, uses an autoregressive (AR) approach to model the distribution of waveform sampling points, but it has to generate a waveform in a time-consuming sequential manner. 
Some new models that use inverse-autoregressive flow (IAF) can generate a whole waveform in a one-shot manner but require either a larger amount of training time or a complicated model architecture plus a blend of training criteria. 

As an alternative to AR and IAF-based frameworks, we propose a neural source-filter (NSF) waveform modeling framework that is straightforward to train and fast to generate waveforms. This framework requires three components to generate waveforms: a source module that generates a sine-based signal as excitation, a non-AR dilated-convolution-based filter module that transforms the excitation into a waveform, and a conditional module that pre-processes the input acoustic features for the source and filter modules. This framework minimizes spectral-amplitude distances for model training, which can be efficiently implemented using short-time Fourier transform routines. As an initial NSF study, we designed three NSF models under the proposed framework and compared them with WaveNet using our deep learning toolkit. 
{It was demonstrated that the NSF models generated waveforms at least 100 times faster than our WaveNet-vocoder, and the quality of the synthetic speech from the best NSF model was comparable to that from WaveNet} on a single-speaker Japanese speech corpus.
\end{abstract}


\IEEEpeerreviewmaketitle

\section{Introduction}
Text-to-speech synthesis (TTS), a technology that converts a text string into a speech waveform \cite{PaulTaylor}, has rapidly advanced 
with the help of deep learning. In TTS systems based on a statistical parametric speech synthesis (SPSS) framework \cite{Tokuda2013,ref:Zen09},
deep neural networks have been used to derive linguistic features from text \cite{yao2015sequence}. More breakthroughs have been reported for the acoustic modeling part, for which various types of neural networks have been proposed that 
convert linguistic features into acoustic features such as the fundamental frequency (F0) \cite{wang2018autoregressive,ribeiro2016parallel}, cepstral or spectral features \cite{zen2013statistical,ref:Fan14,wangARRMDN}, and duration \cite{henter2016robust,Chen2017DiscreteDM}.

Deep learning has recently been used to design a vocoder, a component that converts acoustic features into a speech waveform.
{The pioneering model called WaveNet \cite{oord2016wavenet} directly generates a waveform in a sample-by-sample manner given the input  features.} 
It has shown better performance than classical signal-processing-based vocoders for speaker-dependent SPSS systems \cite{wangICASSP2018}. 
Similar AR models such as SampleRNN \cite{mehri2016samplernn} have also performed reasonably well for SPSS \cite{ai2018samplernn}. 
The key idea of these neural waveform models is to implement an autoregressive (AR) probabilistic model that describes the distribution of a current waveform sample conditioned on previous samples. Although AR models can generate high-quality waveforms, their generation speed is slow because they have to generate waveform samples one by one.

Inverse-AR is another approach to neural waveform modeling.
For example, inverse-AR-flow (IAF) \cite{NIPS2016_6581} can be used to transform a noise sequence into a waveform without the sequential generation process. However, each waveform must be sequentially transformed into a noise-like signal during model training \cite{prenger2018waveglow}, which significantly increases the amount of training time. 
Rather than the direct training method, Parallel WaveNet \cite{pmlr-v80-oord18a} and ClariNet \cite{ping2018clarinet} use a pre-trained AR model as a teacher to evaluate the waveforms generated by an IAF student model. Although this training method avoids sequential transformation, it requires a pre-trained WaveNet as the teacher model. Furthermore, additional training criteria have to be incorporated, without which the student model only produces mumbling sounds \cite{pmlr-v80-oord18a,ping2018clarinet}. The blend of disparate training criteria and the complicated knowledge-distilling approach make IAF-based frameworks even less accessible to the TTS community.

In this paper, we propose a neural source-filter (NSF) waveform modeling framework, which is straightforward to implement, fast in generation, and effective in producing high-quality speech waveforms for TTS. 
NSF models designed under this framework have three components:
a source module that produces a sine-based excitation signal, 
a filter module that uses a dilated-convolution-based network to convert the excitation into a waveform, 
and a conditional module that processes the acoustic features for the source and filter modules. 
The NSF models do not rely on AR or IAF approaches and are directly trained by minimizing the spectral amplitude distances between the generated and natural waveforms.

We describe the three specific NSF models designed under our proposed framework: a baseline NSF (b-NSF) model that adopts a WaveNet-like filter module, simplified NSF (s-NSF) model that simplifies the filter module of the baseline NSF model, and harmonic-plus-noise NSF (hn-NSF) model that uses separate source-filter pairs for the harmonic and noise components of waveforms. While we previously introduced the b-NSF model \cite{wang2018neural}, the s-NSF and hn-NSF models are newly introduced in this paper. Among the three NSF models, the hn-NSF model outperformed the other two in a large-scale listening test. 
Compared with WaveNet, the hn-NSF model generated speech waveforms with comparably good quality but much faster.

We review recent neural waveform models in Section~\ref{sec:review} and describe the proposed NSF framework and three NSF models in Section~\ref{sed:proposed}. After explaining the experiments in Section~\ref{sec:experiments}, we draw a conclusion in Section~\ref{sec:conclusion}.

\section{Review of recent neural waveform models}
\label{sec:review}
We consider neural waveform modeling as the task of mapping input acoustic features into an output waveform.
Suppose one utterance has $B$ frames and its acoustic features are encoded as a sequence $\bs{c}_{1:B}=\{\bs{c}_{1}, \cdots,  \bs{c}_B\}$, where $\bs{c}_b\in\mathbb{R}^{D}$ is the feature vector with $D$ dimensions for the $b$-th frame.  
Without loss of generality, we define the target waveform as $\bs{o}_{1:T}=\{o_1, \cdots, o_T\}$, where $o_t\in\mathbb{R}$, and $T$ is the waveform length\footnote{Some waveform models require discretized waveform values, i.e., $o_t\in\mathbb{N}$.}.
We then use $\widehat{o}_{1:T}$ to denote the waveform generated by a neural waveform model.

\subsection{Naive neural waveform model}
\label{seq:review_naive}
Let us first consider a naive neural waveform model.
After upsampling $\bs{c}_{1:B}$ to $\tilde{\bs{c}}_{1:T}$ by replicating each $\bs{c}_b$ multiple times, 
we may use a recurrent or convolution network to learn the mapping from $\tilde{\bs{c}}_{1:T}$ to ${\bs{o}}_{1:T}$. 
Such a model can be trained by minimizing the mean square error (MSE):
\begin{equation}
\bs{\Theta}^{*} = \arg\min_{\bs{\Theta}} \frac{1}{{T}}\sum_{t=1}^{T}({o}_{t} - \widehat{{o}}_{t})^2,
\label{eq:mse}
\end{equation}
where $\widehat{o}_t = \mathcal{F}_{\Theta}(\bs{c}_{1:T}, t)$ is the output of the neural network at the $t$-th time step for the training utterance, and $\bs{\Theta}$ is the network's parameter set. 
During generation, the model can generate a waveform $\widehat{\bs{o}}_{1:\tilde{T}} = \{\widehat{o}_1, \cdots, \widehat{o}_{\tilde{T}}\}$ by setting $\widehat{o}_t = \mathcal{F}_{\bs{\Theta^*}}(\bs{c}_{1:\tilde{T}}, t), \forall{t}\in\{1,\cdots,\tilde{T}\}$ given the input $\bs{c}_{1:\tilde{T}}$.

However, such a naive model did not produce intelligible waveforms in our pilot test. 
We found that the generated waveform was over-smoothed and lacked the variations that evoke
the perception of speech formant. 
This over-smoothing effect may be reasonable because the way we train the naive model is equivalent to maximizing  
the likelihood of a waveform $\bs{o}_{1:T}$ over the distribution
\begin{equation}
\begin{split}
p(\bs{o}_{1:T} &| \bs{c}_{1:T}; \bs{\Theta})  \\
= &\prod_{t=1}^{T}p({o}_{t} | \bs{c}_{1:T}; \bs{\Theta}) = \prod_{t=1}^{T}\mathcal{N}({o}_{t};  \mathcal{F}_{\bs{\Theta}}(\bs{c}_{1:T}, t), \sigma^2),
\end{split}
\label{eq:naive_model_prob}
\end{equation}
where $\mathcal{N}(\cdot; \cdot , \sigma^2)$ is a Gaussian distribution with an unknown standard deviation $\sigma$.
It is assumed that the waveform values are independently distributed with the naive model, which
may be incompatible with the strong temporal correlation of natural waveforms. 
This mismatch between model assumption and natural data distribution
may cause the over-smoothing effect.


\subsection{Neural AR waveform models}
\label{sec:proposed_model}

AR neural waveform models have recently been proposed for better waveform modeling \cite{oord2016wavenet}.
Contrary to the naive model assumption in Equation~(\ref{eq:naive_model_prob}), it is assumed that
\begin{equation}
p(\bs{o}_{1:T} | \bs{c}_{1:B}; \bs{\Theta})  = \prod_{t=1}^{T}p({o}_{t} | \bs{o}_{<t}, \bs{c}_{1:B}; \bs{\Theta})
\label{eq:ar_model_prob}
\end{equation}
with an AR model,
where $p({o}_{t} | \bs{o}_{<t}, \bs{c}_{1:B}; \bs{\Theta})$ depends on the previous waveform values $\bs{o}_{<t}$. 
Such a model can potentially describe the causal temporal correlation among waveform samples. 

An AR model can be implemented by feeding the waveform sample of the previous time step as the input to a convolution network (CNN) or recurrent network (RNN) at the current step, which is illustrated in Figure~\ref{fig:system_overall}. The model is trained by maximizing the likelihood over natural data while using the data for feedback, i.e., teacher forcing \cite{williams1989learning}. In the generation stage, the model has to sequentially generate a waveform and use the previously generated sample as the feedback datum. 

Pioneering neural AR waveform models include WaveNet \cite{oord2016wavenet} and SampleRNN \cite{mehri2016samplernn}.
There are also models that combine WaveNet with classical speech-modeling methods such as glottal waveform modeling \cite{juvela2018speaker} and linear-prediction coding \cite{valin2018lpcnet,hwang2018lp}.
Although these models can generate high-quality waveforms, their sequential generation process is time-consuming.
{More specifically, these AR models' time complexity to generate a single waveform of length $T$ is theoretically equal to $\mathcal{O}(T)$. Note that the time complexity only takes the waveform length into account, ignoring the number of hidden layers and implementation tricks such as subscale dependency \cite{pmlr-v80-kalchbrenner18a} and squeezing \cite{prenger2018waveglow}}.
Other models, such as WaveRNN \cite{pmlr-v80-kalchbrenner18a}, FFTNet \cite{jin2018fftnet}, and subband WaveNet \cite{okamoto2018investigation}, use different network architectures to reduce or parallelize the computation load, but their generation time is still linearly proportional to the waveform length. 

{Note that in the teacher-forcing-based training stage, the CNN-based (e.g., WaveNet) and RNN-based AR models (e.g., WaveRNN) have time complexity of $\mathcal{O}(1)$ and $\mathcal{O}(T)$, respectively. The time complexity of the RNN-based models is limited by the computation in the recurrent layers. These theoretical interpretations are summarized in Table~\ref{tab:theoretical_compare}.}

\subsection{{IAF-flow-based models}}

{Rather than the AR-generation process, an IAF-flow-based model such as WaveGlow \cite{prenger2018waveglow} uses an invertible and non-AR function $H^{-1}_{\bs{\Theta}}(\cdot)$ to convert a random Gaussian noise signal $\widehat{\bs{z}}_{1:T}$ into a waveform $\widehat{\bs{o}}_{1:T} = H^{-1}_{\bs{\Theta}}(\widehat{\bs{z}}_{1:T},\bs{c}_{1:B})$. 
In the training stage, the model likelihood has to be evaluated as}
\begin{equation}
\begin{split}
&p_{O}(\bs{o}_{1:T} | \bs{c}_{1:B};\bs{\Theta}) \\
= &p_{Z}(\bs{z}_{1:T}=H_{\bs{\Theta}}(\bs{o}_{1:T}, \bs{c}_{1:B}))\Big|\text{det}\frac{\partial{H_{\bs{\Theta}}(\bs{o}_{1:T}, \bs{c}_{1:B})}}{\partial{\bs{o}_{1:T}}}\Big| \\
\label{eq:flow_model}
\end{split}
\end{equation}
where $H_{\bs{\Theta}}(\bs{o}_{1:T}, \bs{c}_{1:B})$ sequentially inverts a training waveform $\bs{o}_{1:T}$ into a noise signal $\bs{z}_{1:T}$ for likelihood evaluation. 
{When $H^{-1}_{\bs{\Theta}}(\cdot)$ is sufficiently complex, $p_{O}(\bs{o}_{1:T} | \bs{c}_{1:B};\bs{\Theta}) $ can approximate the true distribution of $\bs{o}_{1:T}$ \cite{rezende2015variational}.}

{In terms of the theoretical time complexity w.r.t $T$, the flow-based models are dual to the CNN-based AR models, as Table~\ref{tab:theoretical_compare} summaries. Although the time complexity of the flow-based models in waveform generation is irrelevant to $T$, their time complexity in model training is $\mathcal{O}(T)$. Some models such as WaveGlow reduce $T$ by squeezing multiple waveform samples into one vector \cite{prenger2018waveglow} but still require a huge amount of training time.
For example, in one study using Japanese speech data \cite{okamoto-waveglow}, a WaveGlow model was trained on four V100 GPU cards for one month to produce high-quality waveforms\footnote{The original WaveGlow paper used eight GV100 GPU cards for model training \cite{prenger2018waveglow}. However, it did not report the amount of training time.}. }


\subsection{{AR plus IAF}}
Some models, such as Parallel WaveNet \cite{pmlr-v80-oord18a} and ClariNet \cite{ping2018clarinet}, 
combine the advantages of AR and IAF-flow-based models under a distilling framework. 
In this framework, a flow-based student model generates waveform samples $\widehat{\bs{o}}_{1:T} = H^{-1}_{\bs{\Theta}}(\widehat{\bs{z}}_{1:T},\bs{c}_{1:B})$ and queries an AR teacher model to evaluate $\widehat{\bs{o}}_{1:T}$. The student model learns by minimizing the distance between $p(\widehat{\bs{o}}_{1:T})$ and that given by the teacher model. Therefore, neither training nor generation requires sequential transformation.

However, it was reported that knowledge distilling is insufficient and additional spectral-domain criteria must be used \cite{pmlr-v80-oord18a, ping2018clarinet}. 
Among the blend of disparate loss functions, it remains unclear which one is essential. 
Furthermore, knowledge distilling with two large models is complicated in implementation. 


\begin{figure}[t]
\centering
\includegraphics[width=0.95\columnwidth]{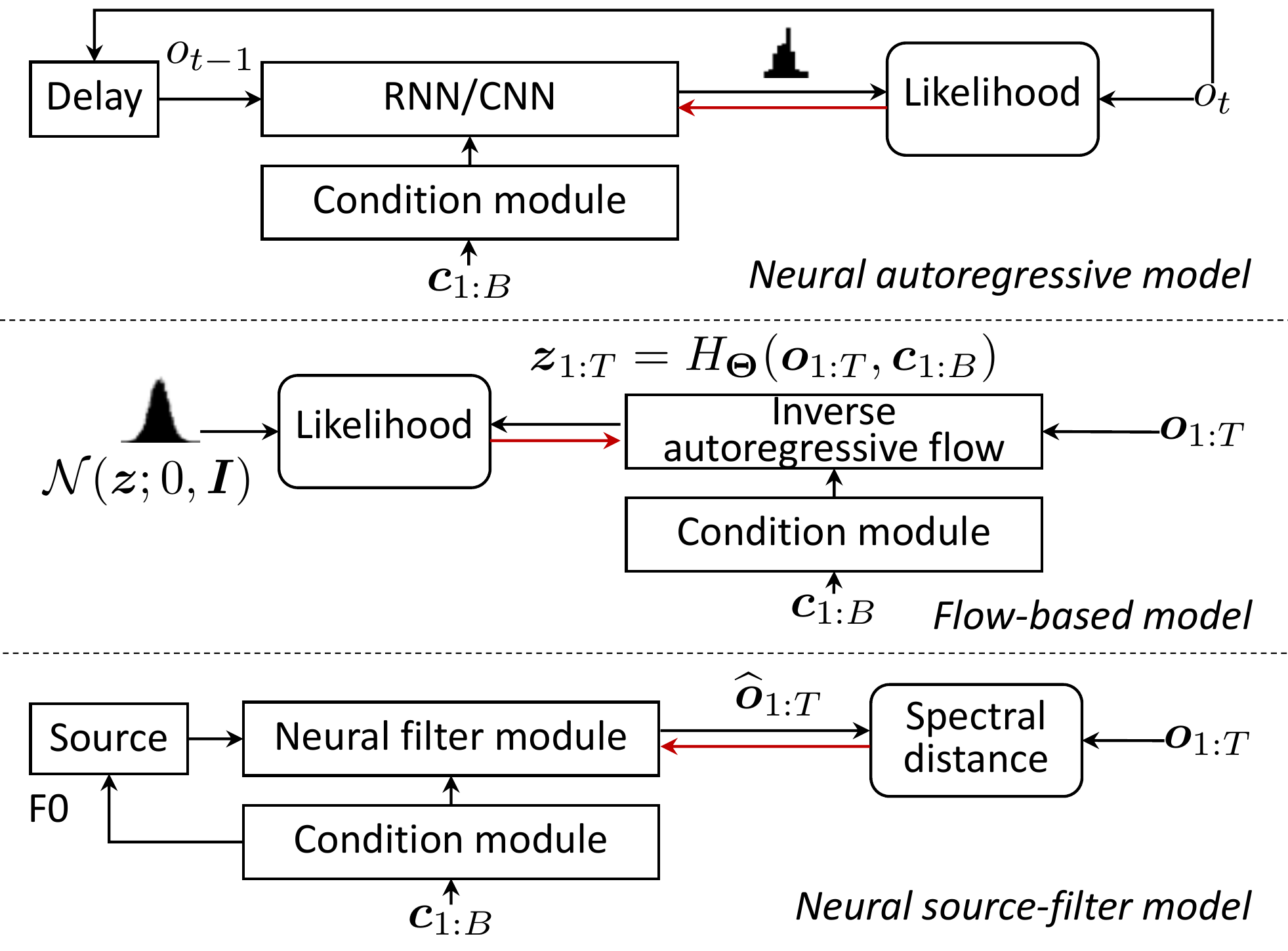}
\vspace{-3mm}
\caption{Three types of neural waveform models in training stage. $\widehat{\bs{o}}_{1:T}$ and $\bs{o}_{1:T}$ denote generated 
and natural waveforms, respectively. $\bs{c}_{1:B}$ denotes input acoustic features. Red arrows denote gradients for back propagation.}
\label{fig:system_overall}
\end{figure}

\section{Neural source-filter models}
\label{sed:proposed}
We propose a framework of neural waveform modeling that is fast in generation and straightforward in implementation.
As Figure~\ref{fig:system_overall} illustrates, our proposed framework contains three modules: 
a condition module that processes input acoustic features, 
source module that produces a sine-based excitation signal given the F0,
and neural filter module that converts the excitation into a waveform using dilated convolution (CONV) and feedforward transformation.
Rather than the MSE or the likelihood over waveform sampling points,
the proposed framework uses spectral-domain distances for model training.  
Because our proposed framework explicitly uses a source-filter structure, we refer to all of the models
based on the framework as NSF models. 

Our proposed NSF framework does not rely on AR or IAF approaches. 
{An NSF model converts an excitation signal into an output waveform without sequential transformation.  
It can be simply trained using the stochastic gradient descent (SGD) method under a spectral domain criterion. 
Therefore, its time complexity is theoretically irrelevant to the waveform length, i.e., $\mathcal{O}(1)$ in both the model training and waveform generation stages. Neither does an NSF model use knowledge distilling, which makes it straightforward in training and implementation.}


%

\begin{table}[!t]
\caption{{Theoretical time complexity to process waveform of length $T$}}
\begin{center}


\begin{tabular}{lccc}
\hline\hline
 		& \multicolumn{2}{c}{Time complexity} & Knowledge \\
\cline{2-3}
Model    	& train & generation &  distilling \\
\hline
AR (CNN-based, e.g., WaveNet)    & $\mathcal{O}(1)$ & $\mathcal{O}(T)$ & $\times$ \\
AR (RNN-based, e.g., WaveRNN)    & $\mathcal{O}(T)$ & $\mathcal{O}(T)$ & $\times$ \\
IAF flow (e.g., WaveGlow)  & $\mathcal{O}(T)$ & $\mathcal{O}(1)$ & $\times$ \\
AR+IAF (e.g., Parallel WaveNet) & $\mathcal{O}(1)$ & $\mathcal{O}(1)$ & \checkmark \\
NSF &  $\mathcal{O}(1)$ & $\mathcal{O}(1)$ & $\times$ \\
\hline\hline
\end{tabular}

\end{center}
\label{tab:theoretical_compare}
\end{table}%

An NSF model can be implemented in various architectures. 
We gives details on the three specific NSF models:
\begin{enumerate} 
\item The b-NSF model has a network structure partially similar to ClariNet and Parallel WaveNet;
\item The s-NSF model inherits b-NSF's structure but uses much simpler neural filter modules;
\item The hn-NSF model extends the s-NSF and explicitly generates the harmonic and noise components of a waveform. 
\end{enumerate}
Although the three NSF models use different neural filter modules, they use the same spectral-domain training criterion. 
We therefore first explain the training criterion in Section~\ref{seq:fft_criterion} 
then the three NSF models in Sections~\ref{sec:model_o_nsf} to \ref{sec:model_h_NSF}.


\subsection{Training criterion based on spectral amplitude distances}
\label{seq:fft_criterion}
A good metric $\mathcal{L}(\widehat{\bs{o}}_{1:T},{\bs{o}}_{1:T})\in\mathbb{R}_{\geq{0}}$ should measure the distance between the perceptual quality of a generated waveform $\widehat{\bs{o}}_{1:T}$ and that of a natural waveform ${\bs{o}}_{1:T}$. 
Additionally, a gradient $\frac{\partial\mathcal{L}}{\partial\widehat{\bs{o}}_{1:T}}$ based on the metric should be easily calculated so that a model can be trained using the SGD method.
For the NSF models, we use spectral amplitude distances calculated on the basis of short-time Fourier transform (STFT).  Although spectral amplitude distance has been used in classical speech coding methods \cite{griffin1985new,Griffin-multiband,griffin1984signal}, we further compute multiple distances with different short-time analysis configurations, as illustrated in Figure~\ref{fig:fig_fft}. 

\subsubsection{\textbf{Evaluation}}
Given the natural and generated waveforms, we follow the STFT convention and conduct framing and windowing on the waveforms, after which the spectrum of each frame is computed using the discrete Fourier transform (DFT). 
Let us use $\widehat{\bs{x}}^{(n)}=[\widehat{x}^{(n)}_1, \cdots, \widehat{x}^{(n)}_M]^{\top}\in\mathbb{R}^{M}$ to denote the $n$-th frame of the generated waveform  $\widehat{\bs{o}}_{1:T}$, where $M$ is the frame length, and $\widehat{x}^{(n)}_m$ is the value of a windowed waveform sample. 
We then use $\widehat{\bs{y}}^{(n)}=[\widehat{y}^{(n)}_1, \cdots, \widehat{y}^{(n)}_K]^{\top}\in\mathbb{C}^{K}$ to
denote the spectrum of $\widehat{\bs{x}}^{(n)}$ calculated using $K$-point DFT.
We similarly define $\bs{x}^{(n)}$ and $\bs{y}^{(n)}$ for the natural waveform $\bs{o}_{1:T}$.
Suppose $\widehat{\bs{o}}_{1:T}$ and ${\bs{o}}_{1:T}$ have been sliced into $N$ frames.  
A log spectral amplitude distance $\mathcal{L}$ can be calculated as
\begin{equation}
\mathcal{L} = \frac{1}{2NK}\sum_{n=1}^{N}\sum_{k=1}^{K}\Big[\log\frac{\texttt{Re}(y_k^{(n)})^2+\texttt{Im}(y_k^{(n)})^2+\eta}{\texttt{Re}(\widehat{y}_k^{(n)})^2+\texttt{Im}(\widehat{y}_k^{(n)})^2+\eta}\Big]^2,
\label{eq:dft_spectral}
\end{equation}
where $\texttt{Re}(\cdot)$ and $\texttt{Im}(\cdot)$ denote the real and imaginary parts of a complex number, respectively, and {$\eta=1\mathrm{e}{-5}$ is a constant number to ensure numerical stability}.

\subsubsection{\textbf{Backward propagation}}
For model training, we need to calculate the gradient vector $\frac{\partial{\mathcal{L}}}{\partial{\widehat{\bs{o}}_{1:T}}}\in\mathbb{R}^{T}$ and propagate it back to the neural filter module that produces $\widehat{\bs{o}}_{1:T}$.
Although $\mathcal{L}$ is calculated given complex-valued spectra, it turns out that $\frac{\partial{\mathcal{L}}}{\partial{\widehat{\bs{o}}_{1:T}}}$ can be computed efficiently.

First, because $\widehat{\bs{y}}^{(n)}=[\widehat{y}^{(n)}_1, \cdots, \widehat{y}^{(n)}_K]^{\top}\in\mathbb{C}^{K}$ is the spectrum of $\widehat{\bs{x}}^{(n)}=[\widehat{x}^{(n)}_1, \cdots, \widehat{x}^{(n)}_M]^{\top}\in\mathbb{R}^{M}$, we know that
\begin{align}
\texttt{Re}(\widehat{{y}}^{(n)}_k) &= {}\sum_{m=1}^{M} \widehat{x}^{(n)}_m \cos\big(\frac{2\pi}{K}(k-1)(m-1)\big), \label{eq:dft_re}\\
\texttt{Im}(\widehat{{y}}^{(n)}_k) &= -\sum_{m=1}^{M} \widehat{x}^{(n)}_m \sin\big(\frac{2\pi}{K}(k-1)(m-1)\big), \label{eq:dft_im}
\end{align}
where $k\in\{1, \cdots,K\}$\footnote{The summation in Equations~(\ref{eq:dft_re})-(\ref{eq:dft_im}) should be $\sum_{m=1}^{K}$, 
but the zero-padded part $\sum_{m=M+1}^{K}0\cos(\frac{2\pi}{K}(k-1)(m-1))$ can be safely ignored. 
Although we can avoid zero-padding by setting $K=M$, in practice, $K$ is usually the power of 2 to take advantage of the fast Fourier transform (FFT). A waveform frame of length $M$ can be zero-padded to length $K=2^{\ceil{\log_{2}{M}}}$ or longer to increase the frequency resolution.}.
Because $\texttt{Re}(\widehat{{y}}^{(n)}_k)$, $\texttt{Im}(\widehat{{y}}^{(n)}_k)$, $\frac{\partial\mathcal{L}}{\partial\texttt{Re}(\widehat{y}_k^{(n)})}$, and $\frac{\partial\mathcal{L}}{\partial\texttt{Im}(\widehat{y}_k^{(n)})}$  are real-valued numbers, we can compute the gradient by using the chain rule:
\begin{equation}
\begin{split}
\frac{\partial\mathcal{L}}{\partial\widehat{{x}}^{(n)}_m} =& 
\sum_{k=1}^{K}\Big[\frac{\partial\mathcal{L}}{\partial\texttt{Re}(\widehat{y}_k^{(n)})}\frac{\partial\texttt{Re}(\widehat{y}_k^{(n)})}{\partial\widehat{{x}}^{(n)}_m} + \frac{\partial\mathcal{L}}{\partial\texttt{Im}(\widehat{y}_k^{(n)})}\frac{\partial\texttt{Im}(\widehat{y}_k^{(n)})}{\partial\widehat{{x}}^{(n)}_m} \Big] \\
=&\sum_{k=1}^{K}\frac{\partial\mathcal{L}}{\partial\texttt{Re}(\widehat{y}_k^{(n)})}\cos(\frac{2\pi}{K}(k-1)(m-1)) - \\
  &\sum_{k=1}^{K}\frac{\partial\mathcal{L}}{\partial\texttt{Im}(\widehat{y}_k^{(n)})}\sin(\frac{2\pi}{K}(k-1)(m-1)).
\label{eq:apped_dft_spectral_grad_2}
\end{split}
\end{equation}
As long as we can compute $\frac{\partial\mathcal{L}}{\partial\widehat{{x}}^{(n)}_m}$ for each $m$ and $n$, the gradient $\frac{\partial{\mathcal{L}}}{\partial{\widehat{{o}}_{t}}}$ for $t\in\{1, \cdots, T\}$ can be easily accumulated from $\frac{\partial\mathcal{L}}{\partial\widehat{{x}}^{(n)}_m}$ given the relationship between $\bs{o}_t$ and each $\widehat{x}^{(n)}_{m}$ that is determined by the framing and windowing operations. 
$\frac{\partial{\mathcal{L}}}{\partial{\widehat{{o}}_{t}}}$ then is sent to the output layer of the neural filter module for back-propagation and SGD training.

\begin{figure}[t]
\centering
\includegraphics[width=\columnwidth]{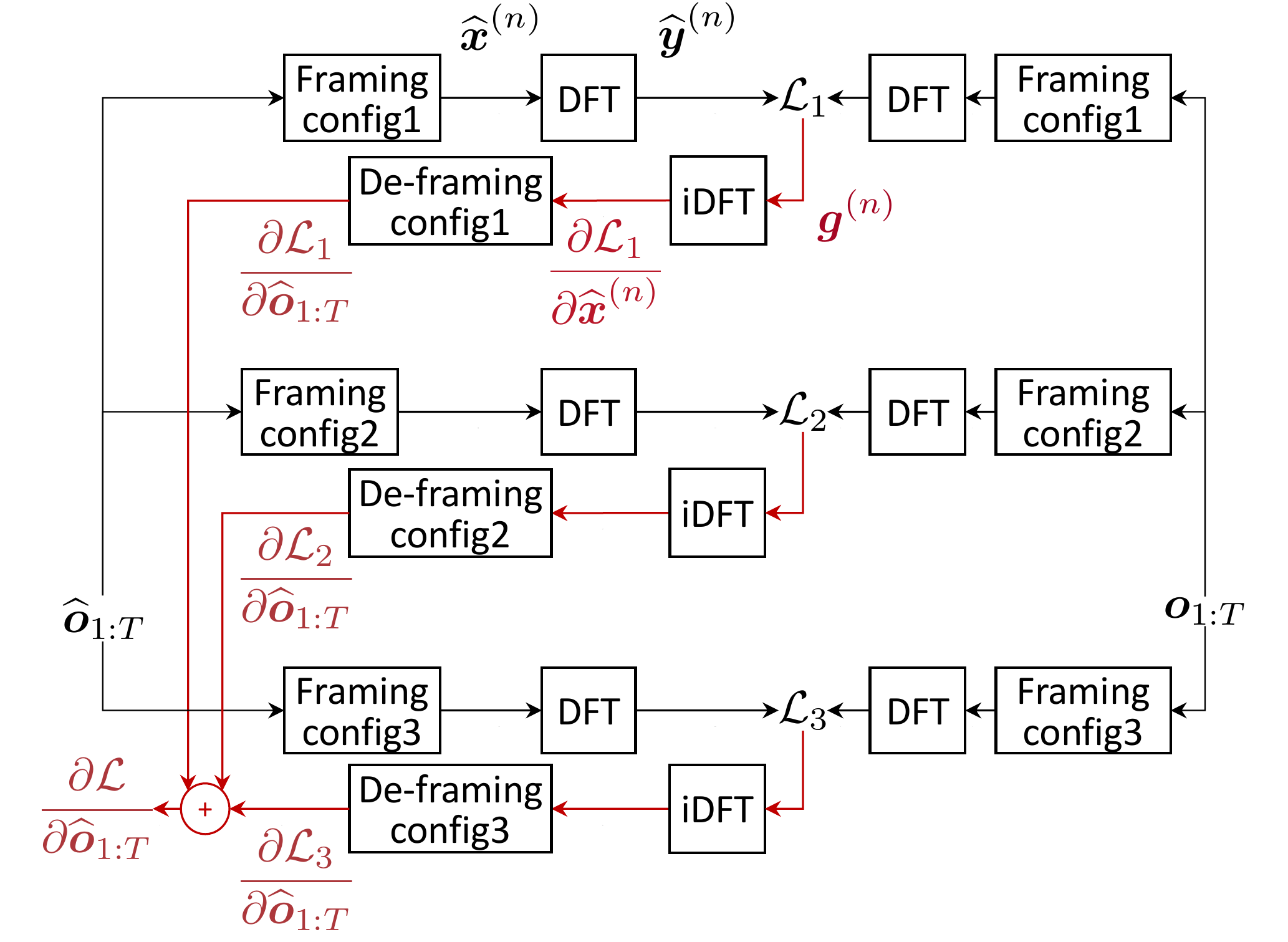}
\vspace{-8mm}
\caption{Illustration of calculating three spectral distances for forward (black) and backward (red) propagation. DFT denotes discrete Fourier transform. Vectors $\widehat{\bs{x}}^{(n)}$, $\widehat{\bs{y}}^{(n)}$, and $\widehat{\bs{g}}^{(n)}$ denote windowed waveform, spectrum, and composed gradient vector for $n$-th frame. }
\label{fig:fig_fft}
\end{figure}

Although we can use matrix multiplication to implement Equation~(\ref{eq:apped_dft_spectral_grad_2}) and compute $\frac{\partial\mathcal{L}}{\partial\widehat{\bs{x}}^{(n)}} = [\frac{\partial\mathcal{L}}{\partial\widehat{{x}}^{(n)}_{1}}, \cdots, \frac{\partial\mathcal{L}}{\partial\widehat{{x}}^{(n)}_{M}}]^{\top}$ for each frame $n$, 
a more efficient method is to use the inverse-DFT.
Suppose a complex-valued signal $\bs{g}^{(n)}\in\mathbb{C}^{K}$ can be composed for the $n$-th frame as $\bs{g}^{(n)} = [{g}^{(n)}_1, \cdots, {g}^{(n)}_K]$, where
\begin{equation}
{g}^{(n)}_k = \frac{\partial{\mathcal{L}}}{\partial{\texttt{Re}(\widehat{{y}}_k^{(n)})}} + j \frac{\partial{\mathcal{L}}}{\partial{\texttt{Im}(\widehat{{y}}_k^{(n)})}}\in\mathbb{C}
\label{eq:g_complex_signal}.
\end{equation}
If $\bs{g}^{(n)}$ is conjugate symmetric, i.e.,
\begin{align}
\texttt{Re}(g_k^{(n)}) &=\texttt{Re}(g_{(K+2-k)}^{(n)}), \quad{}k\in[2, \frac{K}{2}] \label{eq:apped_conjugate_sym_1}\\
\texttt{Im}(g_k^{(n)}) &=
\begin{cases}
-\texttt{Im}(g_{(K+2-k)}^{(n)}), \qquad{}k\in[2, \frac{K}{2}] \\
0, \qquad\qquad\qquad\qquad{}k=\{1, \frac{K}{2}+1\},
\end{cases}
\label{eq:apped_conjugate_sym_2}
\end{align}
the inverse-DFT of $\bs{g}^{(n)}$ will be a real-valued signal $\bs{b}^{(n)}=[b_1^{(n)}, \cdots, b_K^{(n)}]\in\mathbb{R}^{K}$, where
\begin{equation}
\begin{split}
b_m^{(n)} = &\sum_{k=1}^{K} g_k^{(n)} e^{j\frac{2\pi}{K}(k-1)(m-1)}  \\
= &\sum_{k=1}^{K} \texttt{Re}(g_k) \cos(\frac{2\pi}{K}(k-1)(m-1)) -  \\
   & \sum_{k=1}^{K} \texttt{Im}(g_k)\sin(\frac{2\pi}{K}(k-1)(m-1)).
\label{eq:apped_idft_exact}
\end{split}
\end{equation}
By combining Equations~(\ref{eq:g_complex_signal}) and (\ref{eq:apped_idft_exact}), we can 
see that $b_m^{(n)}$ is equal to $\frac{\partial\mathcal{L}}{\partial\widehat{{x}}^{(n)}_m}$ in Equation~(\ref{eq:apped_dft_spectral_grad_2}). 
In other words, if $\bs{g}^{(n)}$ is conjugate symmetric, we can calculate $\frac{\partial\mathcal{L}}{\partial\widehat{\bs{x}}^{(n)}}$ 
by constructing $\bs{g}^{(n)}$ and taking its inverse-DFT. 
It can be shown that $\bs{g}^{(n)}$ is conjugate symmetric when $\mathcal{L}$ is defined as the log spectral amplitude distance in Equation~(\ref{eq:dft_spectral}). Other common distances such as the Kullback-Leibler divergence (KLD) between spectra \cite{lee2001algorithms,Takaki2017} are also applicable. 

Note that, if $\widehat{\bs{x}}^{(n)}$ is zero-padded from length $M$ to length $K$ before DFT, 
the inverse-DFT of $\bs{g}^{(n)}$ will contain gradients w.r.t the zero-padded part. 
In such a case, we can simply assign $\frac{\partial\mathcal{L}}{\partial\widehat{{x}}^{(n)}_m}\leftarrow{b_m^{(n)}}, \forall{m}\in\{1,\cdots,M\}$ and 
ignore $\{b_{M+1}^{(n)}, \cdots, b_K^{(n)}\}$ that correspond to the zero-padded part.

\subsubsection{\textbf{Multi-resolution spectral amplitude distance}}
The previous explanation shows that $\mathcal{L}\in\mathbb{R}$ and $\frac{\partial{\mathcal{L}}}{\partial{\widehat{\bs{o}}_{1:T}}}\in\mathbb{R}^{T}$ can be computed no matter how we configure the window length, frame shift, and DFT bins, i.e., the values of $N$, $M$, $K$.
It is thus straightforward to merge multiple distances $\{\mathcal{L}_{1}, \cdots, \mathcal{L}_{S}\}$ with different windowing and framing configurations. 
In such a case, the ultimate distance can be defined as $\mathcal{L} = \mathcal{L}_{1} + \cdots + \mathcal{L}_{S}$.
Accordingly, the gradients can be merged as $\frac{\partial\mathcal{L}}{\partial{\widehat{\bs{o}}_{1:T}}} = \frac{\partial\mathcal{L}_{1}}{\partial{\widehat{\bs{o}}_{1:T}}} + 
\cdots + \frac{\partial\mathcal{L}_{S}}{\partial{\widehat{\bs{o}}_{1:T}}}$, as illustrated in Figure~\ref{fig:fig_fft}. 

Using multiple spectral distances is expected to help the model learn the spectral details of natural waveforms 
in different spatial and temporal resolutions. We used three distances in this study, which are explained in Section~\ref{sec:model_config}.

\subsubsection{\textbf{Remark on spectral amplitude distance}}
The short-time spectral amplitude distances may be more appropriate than the waveform MSE in Equation~(\ref{eq:mse}) for the NSF models. 
{For a single speech frame, it is assumed with an NSF model using a spectral amplitude distance that 
the spectral amplitude vector $\bs{z} = \bs{y}\odot\bs{y}^{*}\in\mathbb{R}^{K}$ follows a multivariate log-normal distribution with a diagonal covariance matrix, where $\bs{y} = \texttt{DFT}(\bs{x})\in\mathbb{C}^{K}$ is the $K$-point DFT of a waveform frame $\bs{x}$ and $\odot$ denotes element-wise multiplication. 
Although we cannot derive an analytical form of $p(\bs{x})$ from $p(\bs{z})$, we can at least infer that 
the distribution of $\bs{x}$ or the original waveform $\bs{o}_{1:T}$ assumed with the model is at least not an isotropic Gaussian distribution. 
An NSF model with a spectral amplitude distance can potentially model the waveform temporal correlation within an analysis window.}

{Using the spectral amplitude distance is reasonable also because the perception of speech sounds are affected by the spectral acoustic cues such as formants and their transition \cite{fry1962identification,liberman1967perception,strange1989evolving}.
Although the spectral amplitude distance ignores other acoustic cues, such as phase \cite{saratxaga2012perceptual} and timing \cite{hillenbrand1999identification}, we only considered the spectral amplitude distance in this study because we have not found a phase or timing distance that is differentiable and effective.}

\begin{figure*}[h]
\centering
\includegraphics[width=\textwidth]{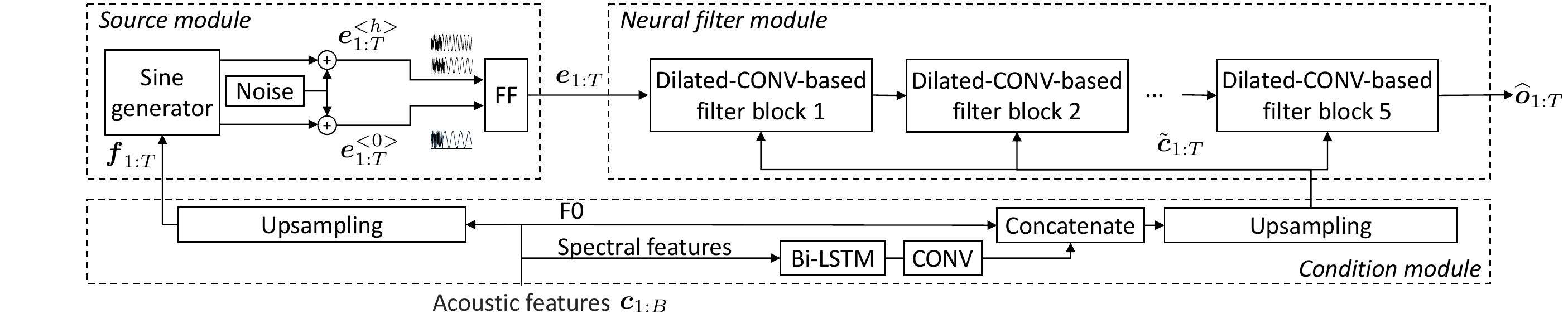}
\vspace{-7mm}
\caption{Structure of baseline NSF (b-NSF) model. $B$ and $T$ denote lengths of input feature sequence and output waveform, respectively. {FF}, {CONV}, and {Bi-LSTM} denote feedforward, convolutional, and bi-directional recurrent layers, respectively. Structure of dilated-CONV filter block is plotted in Figure~\ref{fig:fig_filter_module}.}
\label{fig:fig_o-nsf}
\end{figure*}

\begin{figure*}[h]
\centering
\includegraphics[width=\textwidth]{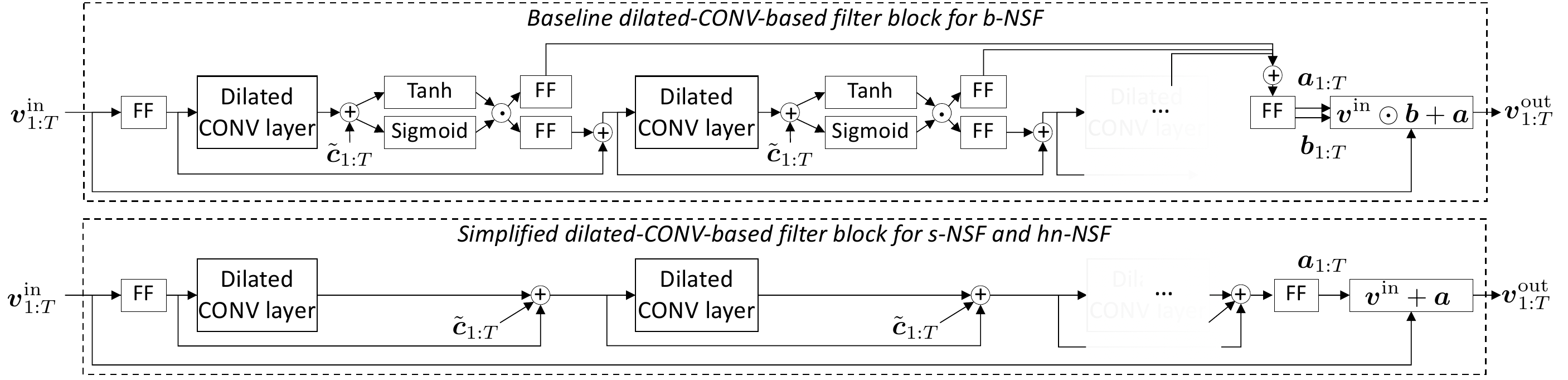}
\vspace{-7mm}
\caption{Structure of dilated-CONV-based filter blocks for b-NSF model (top) and simplified NSF (s-NSF) model (bottom). {$\bs{v}_{1:T}^{\text{in}}$ and $\bs{v}_{1:T}^{\text{out}}$ denote input and output of one filter block, respectively. $\odot$ denotes element-wise product.}
Every filter block contains 10 dilated-convolution layers, and every CONV and FF layer use tanh activation function.}
\label{fig:fig_filter_module}
\end{figure*}

\subsection{Baseline NSF model}
\label{sec:model_o_nsf}
We now give the details on the b-NSF model. 
As Figure~\ref{fig:fig_o-nsf} illustrates, the b-NSF model uses three modules to convert an input acoustic feature sequence $\bs{c}_{1:B}$ of length $B$ into a speech waveform $\widehat{\bs{o}}_{1:T}$ of length $T$:
a source module that generates an excitation signal $\bs{e}_{1:T}$, 
a filter module that transforms $\bs{e}_{1:T}$ into an output waveform, 
and a condition module that processes $\bs{c}_{1:B}$ for the source and filter modules. 
The model is trained using the spectral distance explained in the previous section.

\subsubsection{\textbf{Condition module}}
\label{sec:model_cond}
The input of the condition module is $\bs{c}_{1:B}=\{\bs{c}_1, \cdots, \bs{c}_B\}$, 
where $\bs{c}_{b}=[\tilde{f}_b, \bs{s}_b^{\top}]^{\top}$ includes the F0 value $\tilde{f}_b$ and the spectral features $\bs{s}_b$ for the $b$-th frame. 
The F0 sub-sequence $\{\tilde{f}_1, \cdots, \tilde{f}_B\}$ is upsampled to $\bs{f}_{1:T}$ by replicating each $\tilde{f}_b$ for $\ceil{T/B}$ times, after which $\bs{f}_{1:T}$ is fed to the source module. 
The spectral features are processed by a bi-directional recurrent layer with long-short-term 
memory (LSTM) units \cite{Graves2008} and a 1-dimensional CONV layer with a window size of 3. 
The processed spectral features are then concatenated with the F0 and upsampled as $\tilde{\bs{c}}_{1:T}$. 
The layer size of the Bi-LSTM and CONV layers is 64 and 63, respectively. The dimension of $\tilde{\bs{c}}_{t}$ is 64.
Note that there is no golden network structure for the condition module. 
We used the structure in Figure~\ref{fig:fig_o-nsf} because it has been used in our WaveNet-vocoder \cite{weko_185771_1}
\footnote{Our previous NSF model \cite{wang2018neural} had no ``concatenate'' block in the condition module. 
In this study we added the ``concatenate'' block so that our NSF models use exactly the same condition module as our WaveNet-vocoder. }. 

\subsubsection{\textbf{Source module}}
Given the F0, the source module constructs an excitation signal 
on the basis of sine waveforms and random noise.
In voiced segments, the excitation signal is a mixture of sine waveforms whose frequency values are 
determined by F0 and its harmonics. 
In unvoiced regions, the excitation signal is a sequence of Gaussian noise.

\begin{figure}[!t]
\includegraphics[width=\columnwidth]{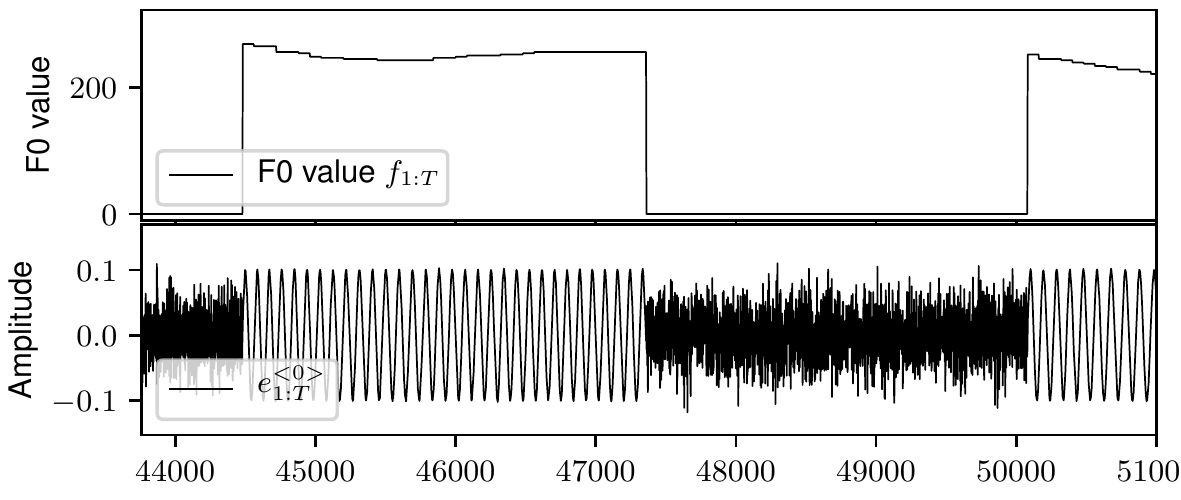}
\vspace{-8mm}
\caption{Example of F0 sequence $\bs{f}_{1:T}$ and fundamental component $\bs{e}_{1:T}^{<0>}$}
\label{fig:f0_sin}
\end{figure}

The input F0 sequence is $\bs{f}_{1:T}$, where $f_t\in\mathbb{R}_{\geq{0}}$ is the F0
value of the $t$-th time step, and $f_t>0$ and $f_t=0$ denote being voiced and unvoiced, respectively. 
A sine waveform $\bs{e}_{1:T}^{<0>}$ with the fundamental frequency can be generated as
\begin{align}
{e}_t^{<0>} = \begin{dcases}
\alpha\sin(\sum_{k=1}^{t}2\pi\frac{{f}_k}{N_s} + {\phi}) + {n}_t, &\text{if }{f_t}>0 \\ 
\frac{\alpha}{3\sigma} {n}_t, & \text{if } f_t = 0\\ 
\end{dcases},
\label{eq:sine}
\end{align}
where ${n}_t \sim \mathcal{N}(0, \sigma^2)$ is Gaussian noise, $\phi\in[-\pi, \pi]$ is a random initial phase, and $N_s$ is the waveform sampling rate. 
The hyper-parameter $\alpha$ adjusts the amplitude of source waveforms, while $\sigma$ is the standard deviation of the Gaussian noise\footnote{In our previous NSF paper \cite{wang2018neural}, we used $\frac{1}{3\sigma}n_t$ for the noise excitation. In this study, we used $\frac{\alpha}{3\sigma}n_t$ so that the amplitude of the noise in unvoiced segments is comparable to that of the sine waveforms in voiced segments.}.
We set $\sigma=0.003$ and $\alpha=0.1$ in this study.
Equation~(\ref{eq:sine}) treats $f_t$ as an instantaneous frequency \cite{carson1937variable}.
Thus, the phase of the $\bs{e}_{1:T}^{<0>}$ becomes continuous even if $f_t$ changes. 
Figure~\ref{fig:f0_sin} plots an example $\bs{e}_{1:T}^{<0>}$ and the corresponding $\bs{f}_{1:T}$.

\begin{figure*}[!t]
\centering
\subfloat{\includegraphics[width=0.5\textwidth]{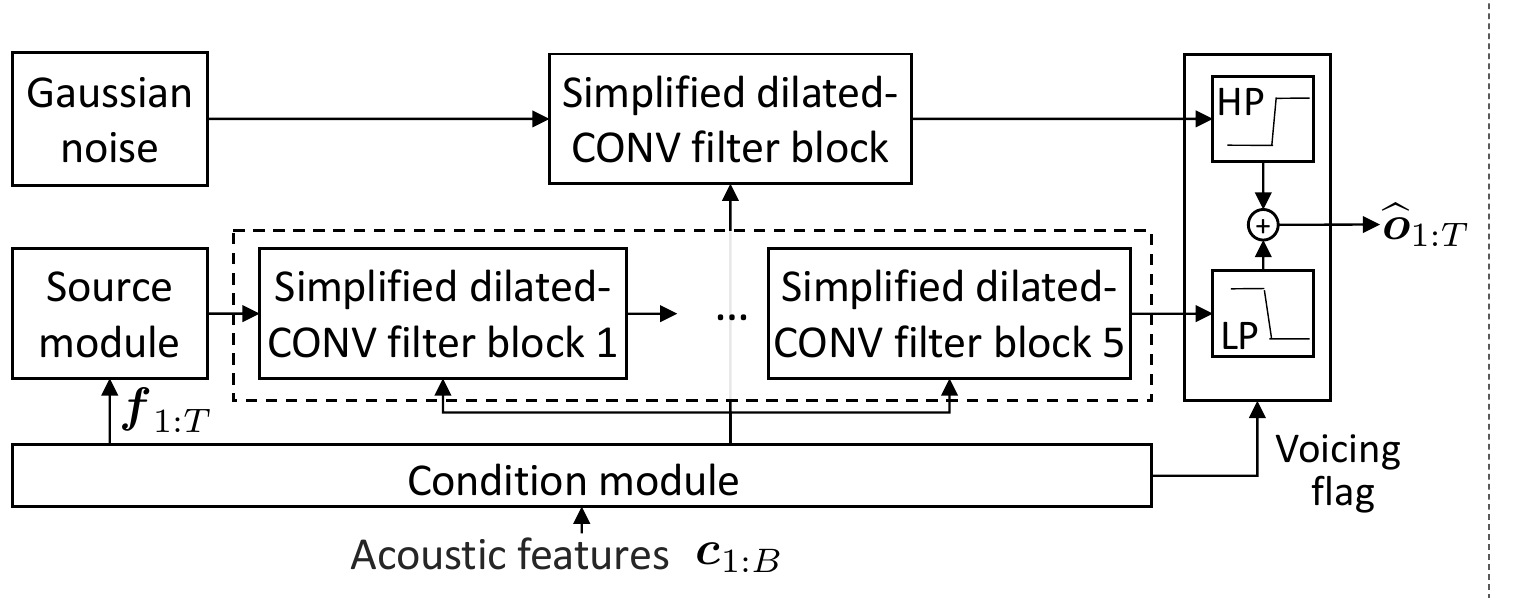}
}
\subfloat{\includegraphics[width=0.48\textwidth]{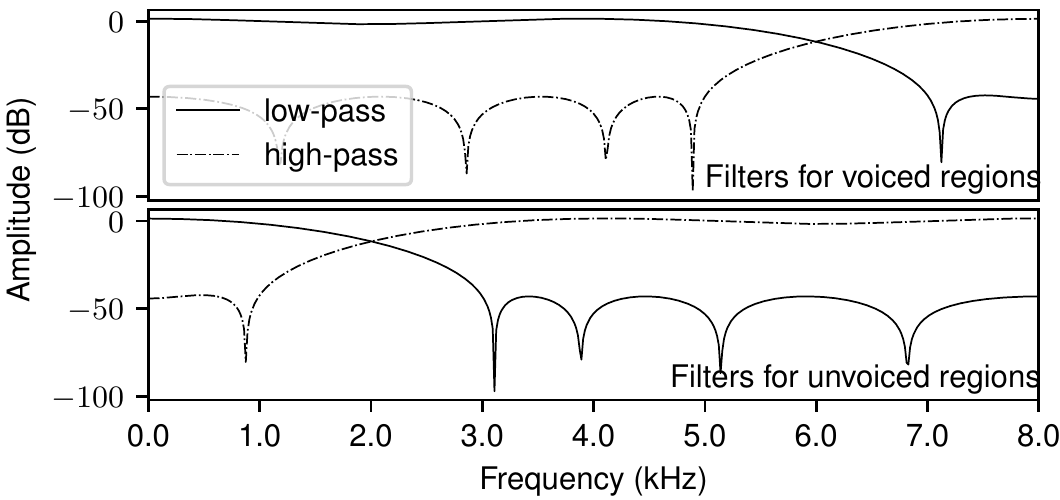}
}
\vspace{-3mm}
\caption{Diagram of harmonic-plus-noise NSF (hn-NSF) model and frequency responses of low-pass (LP) and high-pass (HP) filters for voiced and unvoiced regions.}
\label{fig:i_nsf}
\end{figure*}



{The source module also generates harmonic overtones. 
For the $h$-th harmonic overtone, which corresponds to the $(h+1)$-th harmonic frequency, an excitation signal $\bs{e}_{1:T}^{<h>}$ is generated from Equation~(\ref{eq:sine}) with $(h+1)f_t$. }
The source module then uses a trainable feedforward (FF) layer with a tanh activation function to combine $\bs{e}_{1:T}^{<0>}$ and $\bs{e}_{1:T}^{<h>}$  into the final excitation signal ${\bs{e}_{1:T}}=\{e_1,\cdots, e_T\}$, where ${e}_t\in\mathbb{R}, \forall{t}\in\{1,\cdots, T\}$. This combination can be written as 
\begin{equation}
{\bs{e}_{1:T}}=\text{tanh}(\sum_{h=0}^{H}w_h{\bs{e}_{1:T}}^{<h>} + w_b), 
\label{eq:sum_excitation}
\end{equation}
where $\{w_0, \cdots w_H, w_b\}$ are the FF layer's weights, and $H$ is the total number of overtones. 

{The value of $H$ is not critical to the model's performance because the model can re-create higher harmonic tones, as the experiments discussed in Section~\ref{sec:compare_exp} demonstrated. We set $H=7$ based on a tentative rule $(H+1) * f_{max} < {N_s / 4}$, where $N_s=16$ kHz is the  sampling rate, and $f_{max}\approx{500}$ Hz is the largest F0 value observed in our data corpus. 
We used ${N_s / 4}$ as the upper-bound so that there is at least four sampling points in each period of the sine waveform.}

\subsubsection{\textbf{Neural filter module}}
\label{sec:model_filter}
The filter module of the b-NSF model transforms the excitation ${\bs{e}_{1:T}}$ into an output waveform $\widehat{\bs{o}}_{1:T}$ by using five baseline dilated-CONV filter blocks. The structure of a baseline filter block is illustrated in Figure~\ref{fig:fig_filter_module}.

Suppose the input to the block is $\bs{v}_{1:T}^{\text{in}}$, where $v_t^{\text{in}}\in\mathbb{R}$, $\forall{t}\in\{1,\cdots,T\}$. 
This input $\bs{v}_{1:T}^{\text{in}}$ is expanded in dimension through an FF layer {as $\text{tanh}(\bs{w}{v}_{t}^{\text{in}}+\bs{b})\in{\mathbb{R}^{64}}, \forall{t}\in\{1,\cdots,T\}$, where $w\in{\mathbb{R}}^{64\times{1}}$ is the transformation matrix and $\bs{b}\in{\mathbb{R}^{64}}$ is the bias}.
The expanded signal is then processed by a dilated-CONV layer, summed with the 
condition feature $\tilde{\bs{c}}_{1:T}$, {processed by the gated activation unit based on tanh and sigmoid  \cite{oord2016wavenet}}, and transformed by two additional FF layers.
This procedure is repeated ten times within this filter block, and the dilation size of the dilated convolution layer in the $k$-th stage 
is set to $2^{k-1}$. 
The outputs from the ten stages are summed and transformed into two signals $\bs{a}_{1:T}$ and $\bs{b}_{1:T}$. After that,
${\bs{v}_{1:T}^{\text{in}}}$ is converted into an output signal $\bs{v}_{1:T}^{\text{out}}$ by ${\bs{v}_{1:T}^{\text{out}}}={\bs{v}_{1:T}^{\text{in}}}\odot{\bs{b}_{1:T}} + \bs{a}_{1:T}$, where $\odot$ denotes element-wise multiplication.
The output $\bs{v}_{1:T}^{\text{out}}$ is further processed in the following filter block, and the output of the last filter block is the generated waveform $\widehat{\bs{o}}_{1:T}$.

{Our implementation used a kernel size of 3 for the dilated-CONV layers, which is supposed to be necessary for non-AR waveform models \cite{pmlr-v80-oord18a}.} 
Both the input and output feature vectors of the dilated-CONV layer have 64 dimensions. 
Accordingly, the residual connection that connects two adjacent dilated-CONV layers also has 64 dimensions.
The feature vectors to the FF layer that produces $\bs{a}_{1:T}$ and $\bs{b}_{1:T}$ have 128 dimensions, i.e., skip-connection of 128 dimensions.
The $\bs{b}_{1:T}$ is parameterized as $\bs{b}_{1:T}=\exp(\tilde{\bs{b}}_{1:T})$ to be positive \cite{ping2018clarinet}. 

The baseline dilated-CONV filter block is similar to the student models in ClariNet and Parallel WaveNet because all use the stack of so-called ``dilated residual blocks'' in AR WaveNet \cite{oord2016wavenet}.
However, because the b-NSF model does not use knowledge distilling, it is unnecessary to compute 
the distribution of the signal during forward propagation as ClariNet and Parallel WaveNet do. 
Neither is it necessary to make the filter blocks invertible as IAF does. 
Accordingly, the dilated convolution layers can be non-causal, even though we used causal ones to 
keep configurations of our NSF models consistent with our WaveNet-vocoder in the experiments.

\subsection{Simplified NSF model}
\label{sec:model_s_NSF}
The network structure of the b-NSF model, especially the filter module, was designed on the basis of our experience with implementing WaveNet.
However, we found that the filter module can be simplified, as shown in Figure~\ref{fig:fig_filter_module}. 
Such a filter block keeps only the dilated-CONV layers, skip-connections, and FF layers for dimension change.
The output of a filter block is the sum of a residual signal $\bs{a}_{1:T}$ and the input signal $\bs{v}_{1:T}^{\text{in}}$. 
{Using the sum ${\bs{v}_{1:T}^{\text{out}}}={\bs{v}_{1:T}^{\text{in}}} + \bs{a}_{1:T}$ rather than the affine transformation ${\bs{v}_{1:T}^{\text{out}}}={\bs{v}_{1:T}^{\text{in}}}\odot{\bs{b}_{1:T}} + \bs{a}_{1:T}$ was motivated by the result of our ablation test on the b-NSF model (Section~\ref{sec:ablation_test}: see the results of \texttt{N1}).}
Note that each dilated-CONV layer uses the tanh activation function.

On the basis of the simplified filter block, we constructed the s-NSF model. 
Compared with the b-NSF model, the s-NSF model has the same network structure except the simplified filter blocks.  
Accordingly, the s-NSF model has fewer parameters and a faster generation speed. 
The cascade of simplified filter blocks also turns the neural filter module into a deep residual network \cite{he2016deep}.

\subsection{Harmonic-plus-noise NSF model}
\label{sec:model_h_NSF}
Although the b-NSF and s-NSF models perform well in many cases, 
we found that both may produce low-quality unvoiced sounds or sometimes silence for fricative consonants. 
Analysis on the neural filter modules of a well-trained model suggests that the hidden features for the voiced sounds have a waveform-like
temporal structure, while those for the unvoiced sounds are noisy. 
We thus hypothesize that voiced and unvoiced sounds may require different non-linear transformations in filter modules.

A better strategy may be to generate a periodic and noise component separately and merge them with different ratios into voiced and unvoiced sounds, an idea similar to the harmonic-plus-noise model \cite{abrantes1991hybrid,laroche1993hns,stylianou1996harmonic}.
Following the literature, we also refer to the periodic component as the harmonic component.

As an implementation, we constructed the hn-NSF model, which is illustrated in Figure~\ref{fig:i_nsf}. 
While the hn-NSF model uses the same modules as the s-NSF model to generate a waveform for the harmonic component, it uses only a noise excitation and simplified filter block for the noise component. 
The noise and harmonic components are filtered by a high-pass and low-pass digital filter, respectively, and are summed as the output waveform. Digital filters have been used to merge the noise and periodic signals in the classical speech modeling methods \cite{makhoul1978mixed,mccree1995mixed}. The difference is that the hn-NSF model uses filters to directly merge the waveforms rather than the source signals.

\begin{table}[!t]
\caption{{Parameters of equiripple low- and high-pass FIR filters for hn-NSF. Passband ripple amplitude is less than 5 dB, and stopband  attenuation is -40 dB. Filter coefficients are calculated using Parks-McClellan algorithm \cite{parks1972chebyshev}.}}
\begin{center}
\begin{tabular}{c|cc|cc}
\hline\hline
 & \multicolumn{2}{c|}{Low-pass FIR filter} & \multicolumn{2}{c}{High-pass FIR filter} \\
   & passband & stopband  & passband & stopband\\
\hline
 Voiced sound & $0-5$ kHz & $7-8$ kHz & $7-8$ kHz & $0-5$ kHz  \\
 Unvoiced sound & $0-1$ kHz & $3-8$ kHz & $3-8$ kHz & $0-1$ kHz  \\
\hline\hline
\end{tabular}
\end{center}
\label{tab:filter_design}
\end{table}%


{Because the voiced sounds are usually dominated by the harmonic part, while the unvoiced sounds are dominated by the noise part, we use two pairs of low- and high-pass filters in the hn-NSF model to merge the harmonic and noise components, one pair for voiced sounds and the other for unvoiced sounds.
The configurations of the filters are listed in Table~\ref{tab:filter_design}, }and 
their frequency responses are plotted on the right side of Figure~\ref{fig:i_nsf}. 
We implemented the filters as equiripple finite impulse response (FIR) filters and computed their coefficients by using the Parks-McClellan algorithm \cite{parks1972chebyshev}. 
{Note that the order of the FIR filter is determined by the Parks-McClellan algorithm. For the filters specified in Figure~\ref{fig:i_nsf}, the filter order is around 10}. After the filter coefficients are calculated using the algorithm, they are stored and fixed in the model. The voicing flag for selecting the filter can be easily extracted from the input F0 sequence.

\begin{figure}[!t]
\includegraphics[width=0.95\columnwidth]{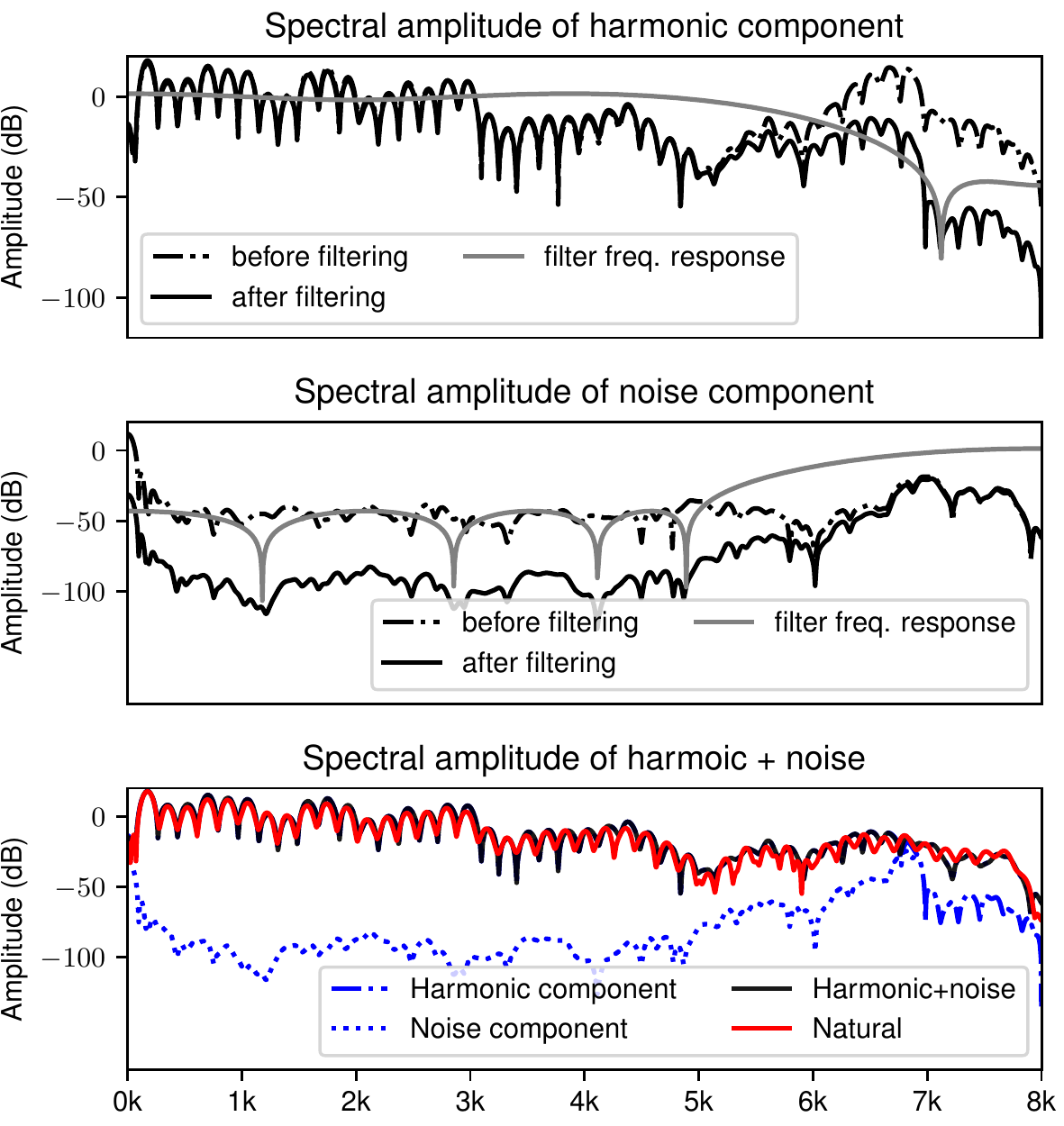}
\vspace{-4mm}
\caption{Spectral amplitude and filter frequency response of one voiced speech frame generated by \texttt{hn-NSF} (Section~\ref{sec:compare_exp}). Note that line of harmonic component overlaps with harmonic+noise and natural one up to 7 kHz.}
\label{fig:filter_response}
\end{figure}

We determined the passband and stopband of the FIR filters after analyzing the spectrogram of the speech data in our corpus.
{Although the passband and stopband are fixed, the neural filter blocks can learn to compensate and fine-tune the energy of the generated signals in certain frequency bands. 
Figure~\ref{fig:filter_response} plots the spectral amplitudes of the harmonic and noise components of one voiced frame generated by the hn-NSF model. For the harmonic component, although the passband of the low-pass filter is only up to 5kHz, the neural filter module generates a harmonic component with a high energy above between 5 and 7 kHz, which compensates for the attenuation of the low-pass filter. As the last row of Figure~\ref{fig:filter_response} shows, the harmonic component dominates the generated waveform frame from 0 to 7 kHz, and its spectral amplitude is similar to that of natural speech. }


\section{Experiments}
\label{sec:experiments}
Following the explanation on our NSF models, we now discuss the experiments. After describing the corpus and data configuration in Section~\ref{sec:corpus_data}, we describe an ablation test conducted on the b-NSF model in Section~\ref{sec:ablation_test}. We then compare the three NSF models with WaveNet in Section~\ref{sec:compare_exp}. In Section~\ref{sec:F0_control}, we investigate the controllability of the input F0 on the NSF models, and in Section~\ref{sec:investigate}, we examine the internal behaviors of the NSF models.

\subsection{Corpus and data configuration}
\label{sec:corpus_data}
Our experiments used a data set of neural-style reading speech by a Japanese female speaker (F009), which is part of the XIMERA speech corpus \cite{kawai2004ximera}. 
This data set was recorded at a sampling rate of 48 kHz and segmented into 30,016 utterances. The total duration is around 50 hours.

We prepared three training subsets for the experiments on waveform modeling: 
the first subset contained 9,000 randomly selected utterances (15 hours), 
the second included 3,000 utterances (5 hours) randomly selected from the first subset, and  
the third contained 1,000 utterances (1.6 hours) randomly selected from the second subset. 
{The first and third subsets were used to evaluate the performance of the NSF models and WaveNet in Section~\ref{sec:compare_exp}. The second subset was used in the ablation test on the NSF models in Section~\ref{sec:ablation_test}.}
We also prepared a validation set with 500 utterances and a test set with another 480 utterances. 
All utterances were downsampled to 16 kHz for waveform model training. 

The acoustic features were extracted from the natural waveforms with a frame shift of 5 ms (200 Hz). 
The F0 values were extracted by an ensemble of pitch trackers \cite{juvelaniibc}. 
Two types of spectral features were prepared: Mel-generalized cepstral coefficients (MGCs) \cite{to1994mel} of order 60 extracted using the WORLD vocoder \cite{morise2016world} 
and Mel-spectrogram of order 80. 
We used the F0 and either the MGCs or the Mel-spectrogram as the input features to the waveform models.

To evaluate the waveform models in SPSS TTS systems, we also trained acoustic models to predict acoustic features
from linguistic features. To train acoustic models, we extracted linguistic features from text, including quin-phone identity, phrase accent type, and other structural information \cite{luong2018investigating}.
These features were force-aligned with the acoustic features by using hidden Markov models.

\subsection{Model configurations}
\label{sec:model_config}
Four waveform models were evaluated in the experiments: our three NSF models \texttt{b-NSF}, \texttt{s-NSF}, and \texttt{hn-NSF} and AR WaveNet (\texttt{WaveNet}). {We chose WaveNet as the benchmark because of its excellent performance reported in both the original paper \cite{oord2016wavenet} and our previous study \cite{wangICASSP2018}\cite{yasuda2019investigation}. We did not include IAF-flow-based models due to their high demand on training time and GPU resources. Neither did we consider Parallel WaveNet or ClariNet due to the lack of authentic implementation.}

\begin{table}[!t]
\caption{Short-time analysis configurations for spectral amplitude distance of NSF models}
\begin{center}
\begin{tabular}{rccc}
\hline\hline
 & $\mathcal{L}_{1}$ & $\mathcal{L}_{2}$ & $\mathcal{L}_{3}$ \\
 \hline
DFT bins $K$         &  512 & 128 & 2048 \\
Frame length $M$  & 320 (20 ms) & 80 (5 ms)  & 1920 (120 ms) \\
Frame shift         & 80 (5 ms)   & 40 (2.5 ms) & 640 (40 ms) \\
\hline\hline
\multicolumn{4}{c}{Note: all configurations used Hann window. } \\
\end{tabular}
\end{center}
\label{tab:dft_config}
\end{table}%

The network configuration of \texttt{b-NSF} was described in Section~\ref{sec:model_o_nsf}. 
It used five baseline dilated-CONV filter blocks, and each block contained ten dilated CONV and other hidden layers, as illustrated in Figure~\ref{fig:fig_filter_module}. The dilation size of the $k$ layer was $2^{k-1}$. 
\texttt{s-NSF} was the same as \texttt{b-NSF} except that each baseline dilated-CONV filter block was replaced with a simplified version. \texttt{hn-NSF} used the same network as \texttt{s-NSF} for the harmonic component and a single simplified block for the noise component. 
{All three NSF models used the spectral amplitude distance $\mathcal{L}=\mathcal{L}_1+\mathcal{L}_2+\mathcal{L}_3$, and the configuration of each $\mathcal{L}_*$ is listed in Table~\ref{tab:dft_config}. Note that $\mathcal{L}_1$ used the same frame shift and frame length as those for extracting the acoustic features from the waveforms. $\mathcal{L}_2$ and $\mathcal{L}_3$ were decided so that one has a higher temporal resolution while the other has a higher frequency resolution than $\mathcal{L}_1$. The configurations in Table~\ref{tab:dft_config} may be inappropriate for a different corpus, and a good configuration may be found through trial and error.}

Our \texttt{WaveNet} used the same network structure as that in our previous study \cite{wangICASSP2018}.
It used 40 dilated-CONV layers in total, where the $k$-th one had a dilation size of $2^{\text{modulo}(k-1,10)}$. 
Between two adjacent CONV layers, \texttt{WaveNet} used a network structure similar to that in the baseline filter block of \texttt{b-NSF}.  
However, each dilated-CONV layer used a kernel size of 2 and 128 output channels, and the condition feature $\tilde{\bs{c}}_{t}$ was transformed by an additional FF layer into 128 dimensions before being summed with the dilated-CONV layer's output.
The skip-connection had 256 dimensions, and the output vectors of the 40 dilated-CONV stages were propagated by the skip-connection and summed
together. The summed vector was transformed into a 1024-dimensional activation vector to a softmax function, which calculates the categorical distribution for the 10-bit quantized $\mu$-law waveform value.
The condition module of \texttt{WaveNet} was the same as the NSF models. 
During generation, \texttt{WaveNet} selected the most probable waveform value from the distribution for 25\% of the voiced time steps. Otherwise, it randomly drew a sample as the output\footnote{This new strategy slightly improved the aperiodicity of voiced sounds, compared with the WaveNet in our previous study, which selected the most probable waveform value at all voiced steps \cite{wangICASSP2018}}.

{All the neural waveform models were trained on a single-GPU card (Nvidia Tesla P100) using the Adam optimizer \cite{kingma2014adam} with 
a learning rate$=0.0003$, $\beta_1=0.9$, $\beta_2=0.999$, and $\epsilon=10^{-8}$. The training process was terminated when the error on the validation set continually increases for five epochs. The batch size was 1, and each utterance was truncated into segments of at most 3 seconds to fit the GPU memory.} 

All the neural waveform models were implemented using a modified CURRENNT toolkit \cite{weninger2015introducing}.  
{Using the same toolkit allows us to fairly compare the models since they use the same set of low-level CUDA/THUST functionalities \cite{bell2011thrust}. Note that our WaveNet implementation is sufficiently good as the benchmark. 
As another study on the same corpus demonstrated \cite{yasuda2019investigation}, our WaveNet implementation can generate speech waveforms that are similar to the original natural waveforms in terms of perceived quality, given natural acoustic features \footnote{Unlike naive open-source implementations, our WaveNet-vocoder avoids redundant CONV operations as the so-called Fast-WaveNet did \cite{paine2016fast}}}. The code, scripts, and samples are publicly available at \url{https://nii-yamagishilab.github.io/samples-nsf/nsf-v2.html}.

For the acoustic models that predict acoustic features from the linguistic features, 
we used shallow and deep neural AR models \cite{wangARRMDN, wang2018autoregressive} to generate the MGCs and F0, respectively. The recipes for training these acoustic models were the same as those in another of our previous study \cite{luong2018investigating}. 
We trained another deep AR model to generate the Mel-spectrogram using a similar recipe. 
The number of training utterances for acoustic models was around 28,000 (47 hours of data).
The acoustic feature sequences were generated given the duration force-aligned on the test set.

\begin{table}[!t]
\caption{Models for ablation test (Section~\ref{sec:ablation_test})}
\begin{center}
\begin{tabular}{cl}
\hline\hline
Model & Description\\
\hline
\texttt{b-NSF} & trained on 5-hr. training set \\
\hline
\texttt{L1} & \texttt{b-NSF} without using $\mathcal{L}_{3}${\textcolor{white}{sssssssssssssssssss}} (i.e., $\mathcal{L} = \mathcal{L}_{1}+\mathcal{L}_{2}$) \\
\texttt{L2} & \texttt{b-NSF} without using $\mathcal{L}_{2}${\textcolor{white}{sssssssssssssssssss}} (i.e., $\mathcal{L} = \mathcal{L}_{1}+\mathcal{L}_{3}$)\\
\texttt{L3} & \texttt{b-NSF} without using $\mathcal{L}_{2}$ or  $\mathcal{L}_{3}${\textcolor{white}{sssssssssssssssss}} (i.e., $\mathcal{L} = \mathcal{L}_{1}$) \\
\hline
\texttt{S1} & \texttt{b-NSF} without harmonics overtones {(i.e., H=0 in Equation~(\ref{eq:sum_excitation}))} \\
\texttt{S2} & \texttt{b-NSF} using noise as source signal {(i.e., $e_{1:T}=\frac{\alpha}{3\sigma}{n_{1:T}}$)} \\
\hline
\texttt{N1} & \texttt{b-NSF} with ${\bs{b}_{1:T}}=1$ in filter blocks \\
\texttt{N2} & \texttt{b-NSF} with ${\bs{b}_{1:T}}=0$ in filter blocks \\
\hline\hline
\end{tabular}
\vspace{-5mm}
\end{center}
\label{tab:models_2}
\end{table}

\subsection{Ablation test on b-NSF}
\label{sec:ablation_test}
Although we claimed that an NSF model can be implemented in varied network architectures,
some of the components may be essential to model performance.
This ablation test was conducted to identify those essential components.

We used \texttt{b-NSF} as the reference model and prepared a few variants, as listed in Table~\ref{tab:models_2}.
All the models including \texttt{b-NSF} were trained using the MGCs, F0, and waveforms from the 5-hr. training set. 
The trained models then generated waveforms given the natural acoustic features in the test set, and these 
generated waveforms were evaluated in a subjective evaluation test. 
In one evaluation round, an evaluator listened to one speech waveform on one screen, rated the speech quality on a 1-to-5 mean-opinion-score (MOS) scale,
and repeated the process for multiple screens.
The waveforms in one evaluation round were for the same text and were played in a random order.
All the waveforms were converted to 16-bit PCM format in advance.

\begin{figure}[!t]
\includegraphics[width=\columnwidth]{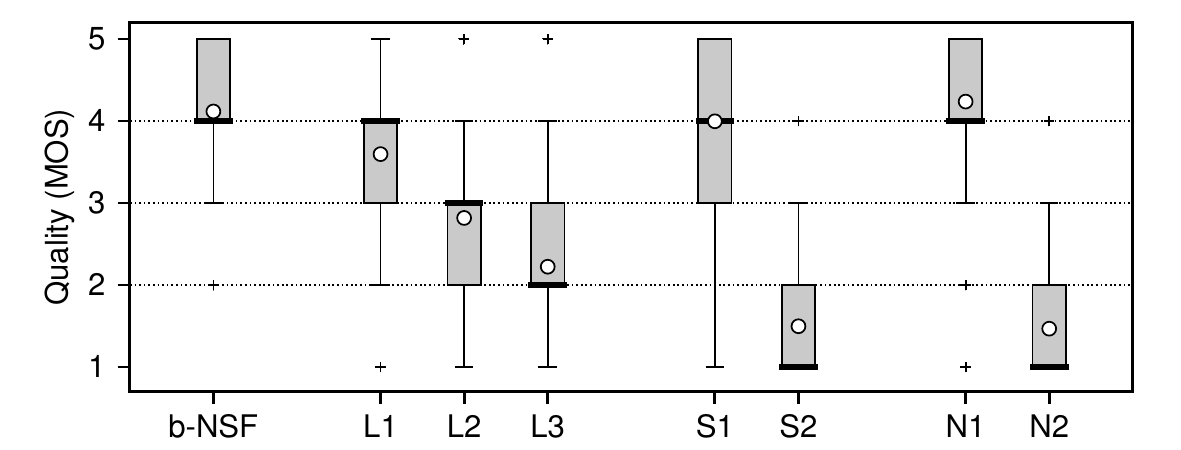}
\vspace{-4mm}
\caption{Mean opinion scores (MOSs) of synthetic samples from \texttt{b-NSF} and its variants given natural acoustic features. White dots denote mean MOSs.}
\label{fig:mos_2}
\end{figure}

\begin{figure}[!t]
\includegraphics[width=\columnwidth]{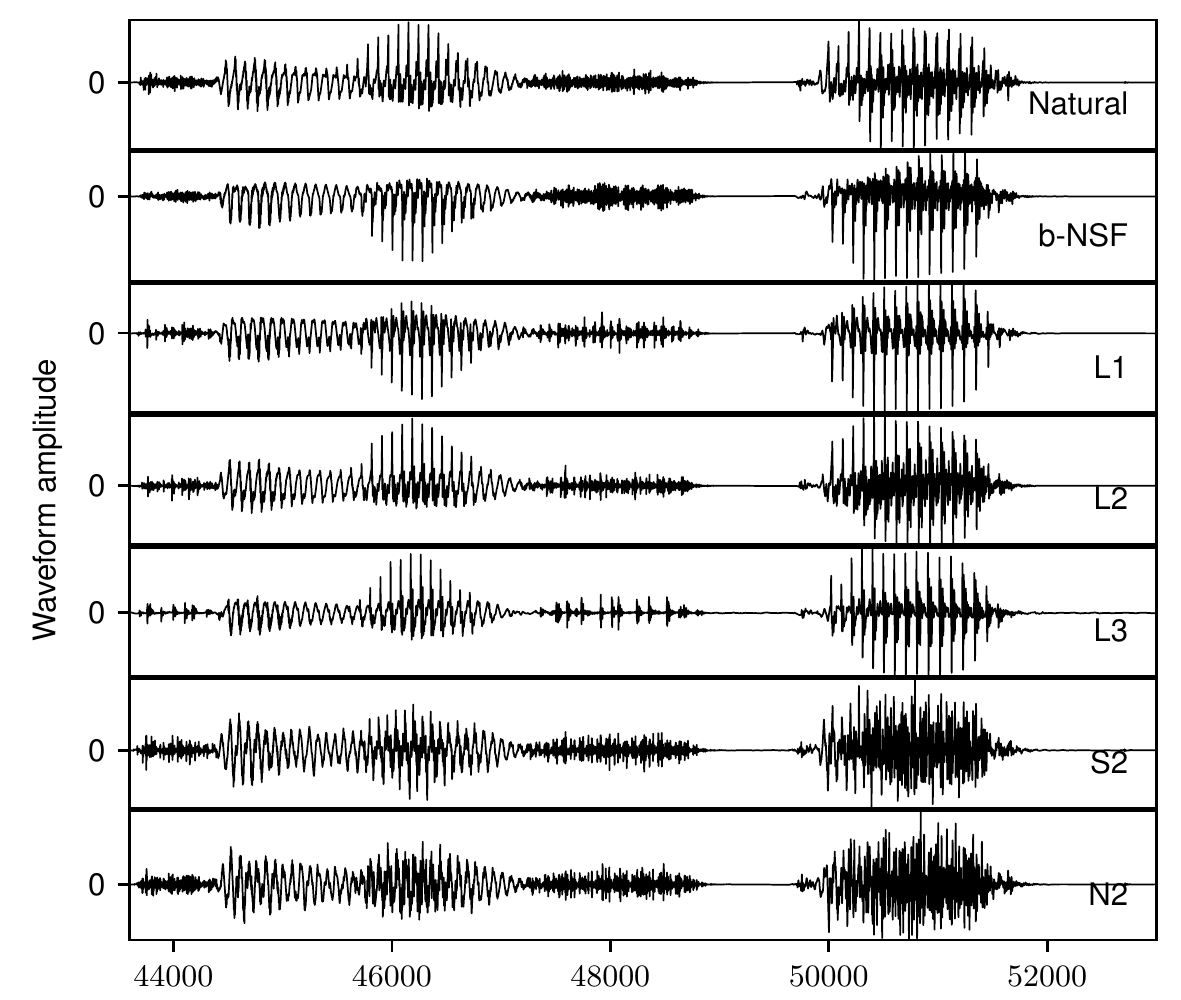}
\vspace{-4mm}
\caption{Natural and generated waveforms from NSF models in ablation test (Section~\ref{sec:ablation_test}).}
\label{fig:waveform_ablation_test}
\end{figure}

\begin{figure*}[h]
\centering
\includegraphics[width=\textwidth]{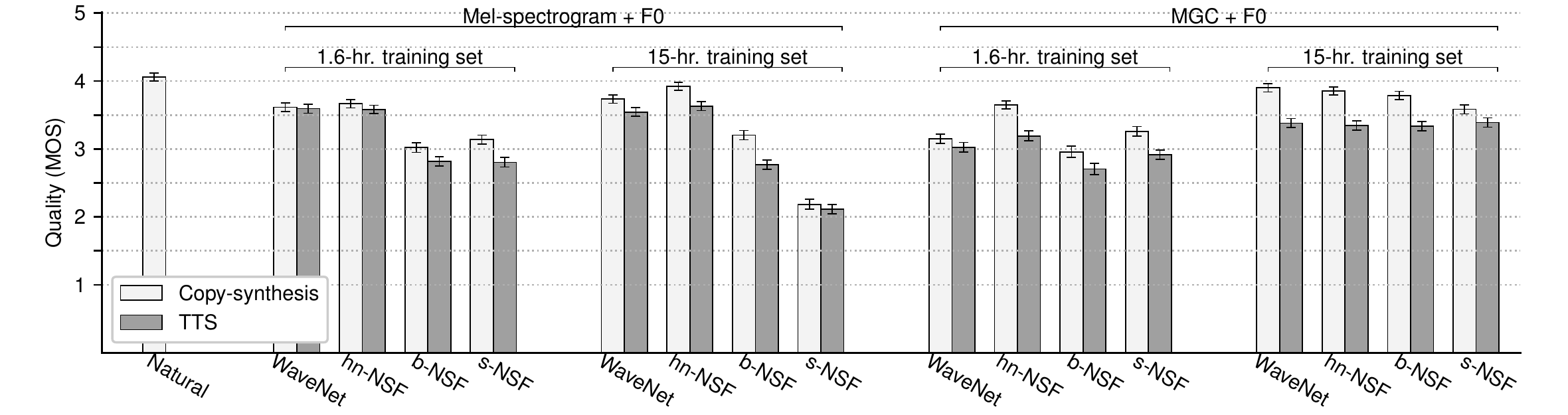}
\vspace{-8mm}
\caption{{MOSs of experimental systems under different training and test conditions. Error bars at confidence level of 95\% are plotted.}}
\label{fig:fig_mos_all}
\end{figure*}

A total of 245 paid Japanese native speakers participated in the test, and 1444 valid evaluation rounds were conducted.
The results are plotted in Figure~\ref{fig:mos_2}, and the difference between \texttt{b-NSF} and the other models was statistically significant ($p<0.01$), as two-sided Mann-Whitney tests demonstrated.
First, a comparison made among \texttt{b-NSF}, \texttt{L1}, \texttt{L2}, and \texttt{L3} showed that using spectral amplitude distances with different windowing and framing configurations is essential to the model's performance. 
The waveforms generated from \texttt{L1}, \texttt{L2}, and \texttt{L3} were perceptually worse because of a pulse-train-like sound in both unvoiced and voiced sounds. This type of artifact could be easily observed in the unvoiced segments plotted in Figure~\ref{fig:waveform_ablation_test}. 
We hypothesize that using spectral distances with different temporal-spatial resolutions could mitigate the pulse-train-like artifacts in the spectrogram.


By comparing \texttt{b-NSF}, \texttt{S1}, and \texttt{S2}, we found that the sine-based excitation is essential to the NSF models. 
In the case of \texttt{S2}, the generated waveforms were intelligible but unnatural because the perceived pitch was unstable. 
As Figure~\ref{fig:waveform_ablation_test} shows, the waveform generated from \texttt{S2} lacked the periodic structure that should be observed in voiced sounds. With a sine-based excitation, the b-NSF model may have a better starting point to generate waveforms with a 
periodic structure. This hypothesis is supported by the results of the investigation discussed in Section~\ref{sec:investigate}.


Interestingly, \texttt{N1} outperformed \texttt{b-NSF} even though it used a simpler transformation in the filter blocks. 
In comparison, the waveforms generated from \texttt{N2} were unnatural because they lacked the stable periodic structures in voiced segments. 
One possible reason is that the skip-connection enables the sine excitation to be propagated to the later filter blocks without being attenuated by
the non-linear transformations.

\begin{figure*}[h]
\centering
\includegraphics[width=\textwidth]{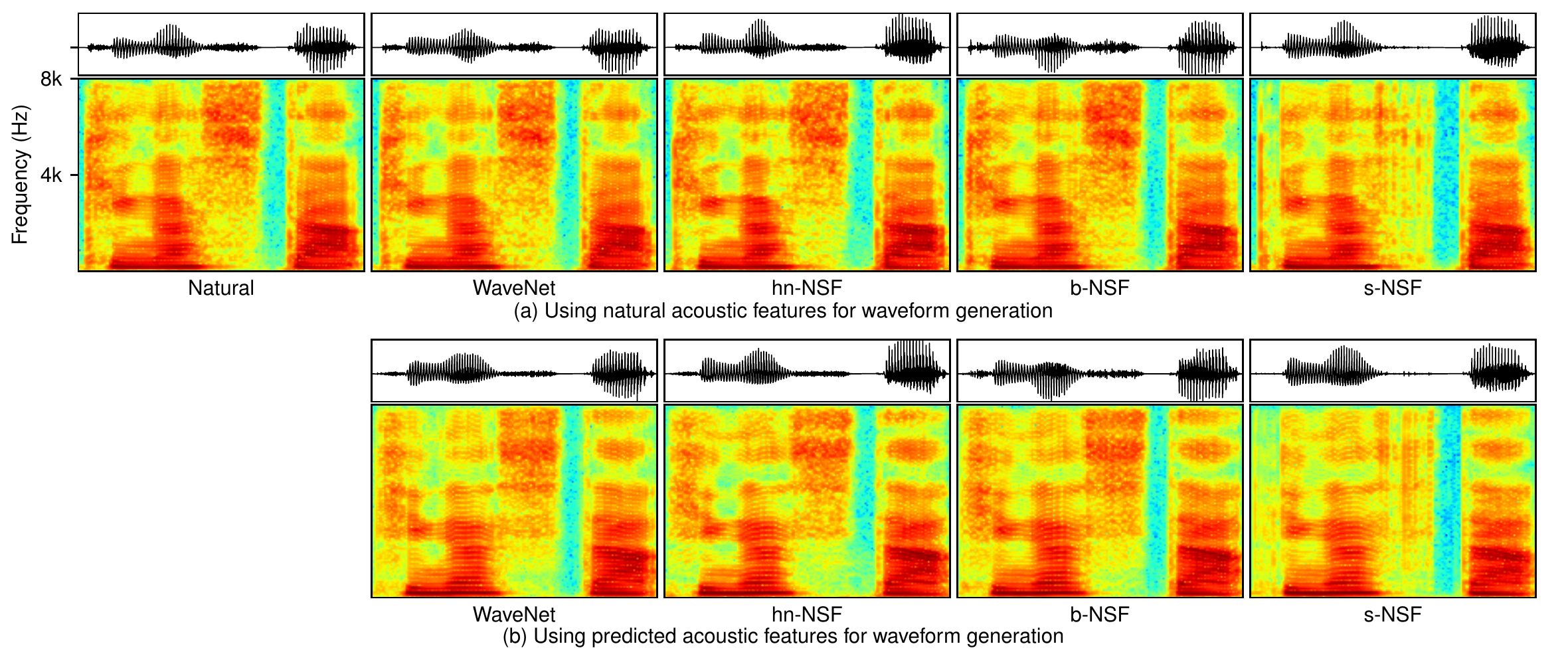}
\vspace{-9mm}
\caption{Natural and generated waveforms and their spectrograms. Models were trained using 15 hours of natural Mel-spectrogram and F0.}
\label{fig:fig_wave_compare}
\end{figure*}

In summary, the results of the ablation test suggest that both the sine excitation and multi-resolution spectral distances are crucial to NSF models.
It is also important to keep the skip-connections inside the filter modules.

\subsection{Comparison between WaveNet and NSF models}
\label{sec:compare_exp}
\subsubsection{\textbf{Quality of generated waveforms}}
This experiment compared \texttt{b-NSF}, \texttt{s-NSF}, \texttt{hn-NSF}, and \texttt{WaveNet} under four training conditions.
Each model was trained using either the 15-hr. or 1.6-hr. training set and conditioned on F0 and either Mel-spectrogram or MGCs. 
In the testing stage, each model generated speech waveforms given natural acoustic features or generated features
produced by the acoustic models. Accordingly, each model was trained and tested under eight conditions.
The generated speech waveforms were evaluated in a subjective evaluation test, which was organized in the same manner as that mentioned in Section~\ref{sec:ablation_test}.

\begin{table}[t!]
\caption{Model comparison in terms of training speed, generation speed, and number of model parameters. Training speed was measured on 15-hr and 1.6-hr training data sets. Training and generation were conducted on single Nvidia P100 GPU card.}
\begin{center}


\setlength{\tabcolsep}{2.8pt}
\begin{tabular}{cccccc}
\hline\hline
 & \multicolumn{2}{c}{Training speed}  & \multicolumn{2}{c}{No. of generated waveforms} & \multicolumn{1}{c}{\multirow[c]{3}{13mm}{No. of model parameters}}  \\
 & \multicolumn{2}{c}{(hr/epoch)}  & \multicolumn{2}{c}{points in 1-s GPU time} & \\
\cline{2-3}\cline{4-5}
Model   & 1.6-hr set & 15-hr set & mem-save & normal &   \\
\hline
 \texttt{WaveNet} & 0.46 & 3.01 &  -   & 0.19 k &  $2.96$e+6 \\
 \texttt{b-NSF}   & 0.37 & 3.40 & 20 k & 227 k  & $1.83$e+6 \\
 \texttt{s-NSF}   & 0.25 & 1.81 & 78 k & 343 k  & $1.07$e+6 \\
 \texttt{hn-NSF}  & 0.27 & 1.92 & 71 k & 335 k  & $1.20$e+6 \\
\hline\hline
\end{tabular}
\end{center}
\label{tab:speed}
\label{tab:size}
\end{table}

The results are plotted in Figure~\ref{fig:fig_mos_all}. 
Among the three NSF models, \texttt{hn-NSF} performed better than or comparably well with the other two versions.
Interestingly, the Mel-spectrogram-based \texttt{s-NSF} performed poorly when it was trained using the 15-hr. set. One reason was that \texttt{s-NSF} produced low-quality unvoiced sounds. For example, the unvoiced segments generated by \texttt{s-NSF} had a very small amplitude, as shown in Figure~\ref{fig:fig_wave_compare}.
\texttt{b-NSF} performed well when it was trained using MGCs and F0 from the 15-hr. training set,
but its performance was worse than \texttt{hn-NSF} and \texttt{b-NSF} when the amount of training data was less than 2 hours. 
One hypothesis is that \texttt{b-NSF} requires more training data since it has more parameters, as Table~\ref{tab:size} shows. 

A comparison made between \texttt{hn-NSF} and \texttt{WaveNet} shows that \texttt{hn-NSF} was comparable to
\texttt{WaveNet} in terms of the generated speech quality. 
Specifically, in the TTS application, \texttt{hn-NSF} trained on 15 hours of Mel-spectrogram data slightly outperformed \texttt{WaveNet} and \texttt{b-NSF} trained on 15 hours of MGC data, which were the best performing models in our previous study \cite{wang2018neural}.

\subsubsection{\textbf{Training and generation speed}}
After the MOS test, we compared the waveform training and generation speed of the experimental models. 
Although the theoretical time complexity is described in Table~\ref{tab:theoretical_compare},  
we measured the actual speed of each model in training and generation stages. 

{Table~\ref{tab:speed} lists the time cost to train the experimental models for one epoch.
The training time cost on the 15-hr training set was larger than that on the 1.6-hr set because the number of training utterance increased. Nevertheless, the results indicate that the NSF models are comparable to \texttt{WaveNet} in terms of training speed. This is expected because all three NSF models and \texttt{WaveNet} require no sequential transformation of waveforms during model training.}

For waveform generation, our implementation of the NSF models has normal and memory-saving generation modes. The normal mode allocates all of the required GPU memory once but cannot generate very long waveforms due to the limited memory space on a single GPU card. The memory-saving mode supports the generation of long waveforms because it releases and allocates the GPU memory layer by layer. However, these memory operations cost processing time. 

We evaluated the NSF models in both modes by using a test subset with 80 test utterances, each of which was around 5 seconds.  As the results in Table~\ref{tab:speed} indicate, the NSF models were much faster than \texttt{WaveNet} even in the memory-save mode. Note that \texttt{WaveNet} requires no repeated memory operation or memory-save mode. It is slow because of the AR-generation process. Compared with \texttt{b-NSF}, \texttt{s-NSF} was faster in generation because of its simplified network structure. Although \texttt{hn-NSF} slightly lagged behind \texttt{s-NSF} in terms of generation speed, it outperformed \texttt{b-NSF}. 

In summary, the results of the MOS and speed tests indicate that \texttt{hn-NSF} performed no worse than \texttt{WaveNet} in terms of the quality of the generated waveforms. Furthermore,  \texttt{hn-NSF} outperformed \texttt{WaveNet} by a large margin in term of waveform generation speed\footnote{{The generation speed in the memory-save mode may be further increased if the GPU memory allocation can be accelerated. 
}}.

\begin{figure*}[h]
\centering
\includegraphics[width=\textwidth]{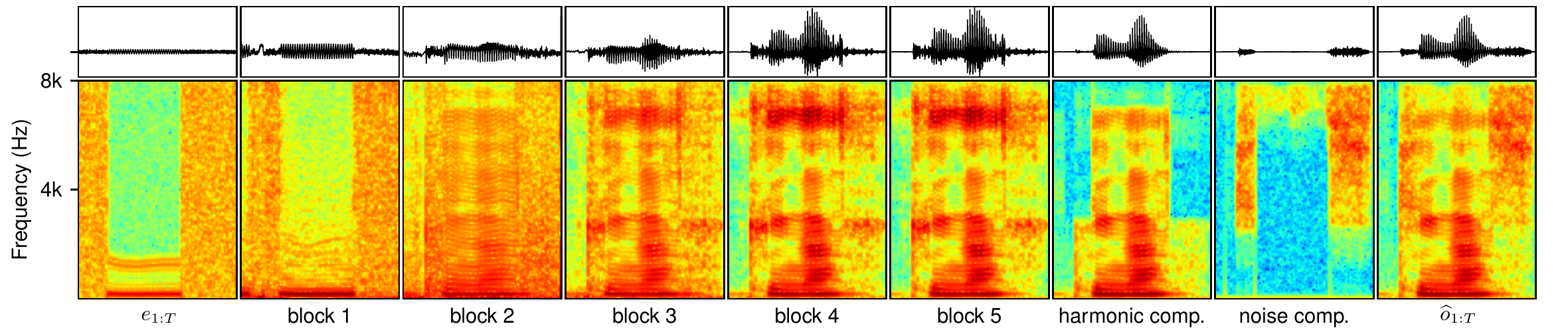}
\vspace{-8mm}
\caption{Input, output, and hidden signals from \texttt{hn-NSF} given natural Mel-spectrogram and F0.
From left to right: input excitation $\bs{e}_{1:T}$, outputs from 5 neural filter blocks for harmonic waveform component, filtered harmonic and noise waveform components, and output waveform $\widehat{\bs{o}}_{1:T}$.}
\label{fig:fig_wave_analysis}
\vspace{-4mm}
\end{figure*}

\subsection{Consistency between input F0 and F0 in generated waveforms}
\label{sec:F0_control}
The ablation test discussed in Section~\ref{sec:ablation_test} demonstrated that sine excitation with the input F0 is essential to the NSF models.
In this experiment, we investigated the consistency between the input F0 and the F0 of the waveforms generated from the NSF models, especially when 
the F0 input to the source module is not identical to the F0 in the input Mel-spectrogram.

This experiment was conducted on \texttt{hn-NSF} and \texttt{WaveNet}, which were trained using the natural Mel-spectrogram and F0 from the 15-hr. training set. The \texttt{WaveNet} was included as a reference model.
Before generating the waveforms, we used the deep AR F0 model to randomly generate three F0 contours for each of the test set utterances \cite{wang2018autoregressive}.
In this random generation mode, the three F0 contours for the same test utterance were slightly different from each other \cite{wang2018autoregressive}.
Let \textbf{F0r1}, \textbf{F0r2}, and \textbf{F0r3} denote the three sets of F0 contours.
We then used the three F0 sets and the generated Mel-spectrogram as the input to \texttt{hn-NSF} and \texttt{WaveNet}, which resulted in six sets of generated waveforms.

For each of the six sets, we calculated the correlation between the input F0 and the F0 extracted from the generated waveforms.
For reference, {we also extracted the F0 from the input Mel-spectrograms using a neural-network-based method \cite{juvela2018speech}}. 
The results listed in Table~\ref{tab:f0_corr} indicate that the F0 contours of the waveforms generated from the NSF models were highly consistent with the F0 input to the source module.  
However, the waveforms generated from WaveNet correlated with the F0 information buried in the input Mel-spectrogram.


The results indicate that we can easily control the F0 of the generated waveforms from the NSF models through directly manipulating the input F0. In contrast, it is less straightforward in the case of WaveNet because we have to manipulate the F0 contained in the input  Mel-spectrogram.

\begin{table}[t!]
\caption{F0 correlation between F0 input to waveform models and F0 extracted from waveforms generated by \texttt{hn-NSF} and \texttt{WaveNet}. Note that \texttt{hn-NSF} and \texttt{WaveNet} used Mel-spectrogram and F0 in \textbf{F0r1}, \textbf{F0r2}, or \textbf{F0r3}.}
\begin{center}
\begin{tabular}{ccccccc}
\hline\hline
	& \multicolumn{3}{c}{\texttt{hn-NSF}} & \multicolumn{3}{c}{\texttt{WaveNet}} \\
	\cline{2-4}\cline{5-7}
        & \textbf{F0r1} & \textbf{F0r2} & \textbf{F0r3} & \textbf{F0r1} & \textbf{F0r2} & \textbf{F0r3} \\
  \hline
  \textbf{F0r1} &  \textbf{0.992} & 0.930 & 0.921 & 0.917 & 0.916 & 0.914\\
  \textbf{F0r2} &  0.926 & \textbf{0.986} & 0.919 & 0.918 & 0.919 & 0.917\\
  \textbf{F0r3} &  0.926 & 0.929 & \textbf{0.988} & 0.922 & 0.921 & 0.920\\
\hline
F0 in Mel-spect. &  0.901 & 0.907 & 0.900 & \textbf{0.975} & \textbf{0.971} & \textbf{0.973} \\
\hline\hline
\end{tabular}
\end{center}
\label{tab:f0_corr}
\end{table}

\subsection{Investigation of hidden features of hn-NSF}
\label{sec:investigate}
We argued in Section~\ref{sec:model_s_NSF} that the simplified neural filter module in the s-NSF and hn-NSF models is similar to a deep residual network. It is thus interesting to look inside the neural filter module.
From $\texttt{hn-NSF}$ trained on 15 hours of Mel-spectrogram and F0, we generated one test utterance given natural condition data and extracted the one-dimensional output of each simplified filter block (i.e., the $\bs{v}_{1:T}^{\text{out}}$ in the bottom panel of Figure~\ref{fig:fig_filter_module}) in the sub-network to generate the harmonic waveform component. 
We also extracted the sine-based excitation $\bs{e}_{1:T}$, filtered harmonic and noise waveform components, and final output waveform $\widehat{\bs{o}}_{1:T}$. These signals and their spectrograms are plotted in Figure~\ref{fig:fig_wave_analysis}. 

We can observe that the dilated-CONV filter blocks morphed the sine excitation into the waveform.
The spectrogram of the sine excitation had no formant structure but only the fundamental frequency and harmonics. 
From blocks 1 to 5, the spectrogram of the signal was gradually enriched with the formant structure. 
The results also suggest that \texttt{hn-NSF} kept the F0 of the sine excitation in the output waveform.
This explains why the F0 of the waveform generated from \texttt{hn-NSF} was highly consistent with the frequency of the sine excitation, or the input F0,  the experiments of Section~\ref{sec:F0_control}.

Similar results were observed when we analyzed \texttt{s-NSF} and \texttt{b-NSF}.
For \texttt{b-NSF}, the results are consistent with the ablation test where we found that
\texttt{b-NSF} without the skip-connections in the filter module performed poorly (\texttt{N2} in Section~\ref{sec:ablation_test}).
The skip-connections make the filter module a deep residual network based on which the excitation signal can be 
gradually transformed into the output waveform. 

These results also indicate how the sine excitation eases the task of waveform modeling because the neural filter modules do not need to reproduce the periodic structure that evokes the perception of F0 in voiced sounds. Without the sine excitation, it may be difficult for the neural filter modules to generate the periodic structure, which explains the poor performance of the b-NSF model without sine excitation (\texttt{S2} in Section~\ref{sec:ablation_test}).

\section{Conclusion}
\label{sec:conclusion}
We proposed a framework called ``neural source-filter modeling'' for the waveform models in TTS systems.
A model implemented in this framework, which is called an ``NSF model'', can convert input acoustic features
into a high-quality speech waveform. Compared with other neural waveform models such as WaveNet, an NSF model does not use an AR network structure and avoids the slow sequential waveform generation process. Neither does an NSF model use flow-based approaches nor knowledge distilling. 
Instead, an NSF model uses three modules that can be easily implemented: a source module that produces a
sine-based excitation signal, filter module that transforms the excitation into an output waveform, and condition
module that processes the input features for the source and filter modules. 
Such an NSF model can be efficiently trained using a merged spectral amplitude distance.
Even though this distance is calculated using multiple short time analysis configurations, it can be efficiently implemented on the basis of STFT. Therefore, the proposed NSF framework allows a neural waveform model to be built and trained straightforwardly.

Experimental results indicated that the specific hn-NSF model, which uses separate modules to model the harmonic and noise components of waveforms, performed comparably well to our WaveNet on a large single-speaker Japanese speech corpus. 
Furthermore, this NSF model can generate speech waveforms at a much faster speed. 
Another advantage of this NSF model is that the F0 input to the source module allows easy control on the pitch of the generated waveform.

{In this primary study, we mainly described the NSF framework in detail and compared several NSF models with our verified WaveNet implementation. To further understand the NSF models' performance, we need a thorough comparison between NSF models and other types of neural waveform models on multi-speaker corpora. We leave this task for future work because of the time required to implement and train other models.}

\ifCLASSOPTIONcaptionsoff
  \newpage
\fi

\bibliographystyle{IEEEtran}
\bibliography{./BIB.bib}

\end{document}